\documentclass{bmcart}

\usepackage{verbatim}
\usepackage{listings}
\usepackage{ulem,xcolor}

\usepackage{fancyvrb}
\usepackage{fvextra}
\usepackage{hyperref}

\startlocaldefs

\usepackage{graphicx}

\newcommand{\IBDorfURL}{\href{https://bitbucket.org/mjponce/racs/src/e5c3be16ee6556b267f75d114a3c0ede216a20fd/datasets/IBD1_ORF.xlsx?at=master}{RACS/datasets/IBD1\_ORF.xlsx}}

\newcommand\redsout{\bgroup\markoverwith{\textcolor{red}{\rule[0.5ex]{2pt}{1.4pt}}}\ULon}

\RecustomVerbatimCommand{\VerbatimInput}{VerbatimInput}{fontsize=\footnotesize,
  frame=lines,   framesep=2em,                          breaklines=true,
 breakanywhere=true
}

\endlocaldefs

\begin{document}

\begin{frontmatter}

\begin{fmbox}
\dochead{Methodology article}

\title{RACS: Rapid Analysis of ChIP-Seq data for contig based genomes}

\author[
   addressref={aff1},
   noteref={n1},
   email={mponce@scinet.utoronto.ca}
]{\inits{MP}\fnm{Marcelo} \snm{Ponce}}

\author[
   addressref={aff2},
   noteref={n1},
   email={alejandro.saettone@ryerson.ca}
]{\inits{AS}\fnm{Alejandro} \snm{Saettone}}

\author[
   addressref={aff3},
   email={nabeel.haidershah@mail.utoronto.ca}
]{\inits{SNS}\fnm{Syed} \snm{Nabeel-Shah}}  

\author[
   addressref={aff2},
   corref={aff2},
   email={jeffrey.fillingham@ryerson.ca}
]{\inits{JF}\fnm{Jeffrey} \snm{Fillingham}}

\address[id=aff1]{
  \orgname{SciNet High Performance Computing Consortium, University of Toronto},
  \street{661 University Ave},
  \postcode{M5G 1M1}
  \city{Toronto},
  \cny{Canada}
}

\address[id=aff2]{
  \orgname{Department of Chemistry and Biology, Ryerson University},
  \street{350 Victoria St},
  \postcode{M5B 2K3}
  \city{Toronto},
  \cny{Canada}
}

\address[id=aff3]{
  \orgname{Department of Molecular Genetics, University of Toronto},
  \street{1 King's College Cir},
  \postcode{M5S 1A8}
  \city{Toronto},
  \cny{Canada}
}

\begin{artnotes}
\note[id=n1]{Equal contributor}
\end{artnotes}

\end{fmbox}

\begin{abstractbox}

\begin{abstract}
\textbf{Background:} 
Chromatin immunoprecipitation coupled to next generation sequencing (ChIP-Seq) is a widely used technique to investigate the function of chromatin-related proteins in a genome-wide manner.
ChIP-Seq generates large quantities of data which can be difficult to process and analyse,
particularly for organisms with contig based genomes.
Contig-based genomes often have poor annotations for
cis-elements, for example enhancers, that are important for gene expression.
Poorly annotated genomes make a comprehensive analysis of ChIP-Seq data
difficult and as such standardized analysis pipelines are lacking.

\textbf{Methods:}
We report a computational pipeline that utilizes
traditional High-Performance Computing techniques
and open source tools for processing and analysing data obtained from ChIP-Seq.

We applied our computational pipeline ``Rapid Analysis of ChIP-Seq data''
(RACS) to ChIP-Seq data that was generated in the model organism
\textit{Tetrahymena thermophila}, an example of an organism with a genome that is
available in contigs.

\textbf{Results:}
To test the performance and efficiency of RACs, we performed control
ChIP-Seq experiments allowing us to rapidly eliminate false positives
when analyzing our previously published data set.
Our pipeline segregates the found read accumulations between genic
and intergenic regions and is highly efficient for rapid downstream analyses.

\textbf{Conclusions:}
Altogether, the computational pipeline presented in this report is an
efficient and highly reliable tool to analyze genome-wide ChIP-Seq data
generated in model organisms with contig-based genomes.
\end{abstract}

\begin{keyword}
\kwd{Chromatin Immunoprecipitation}
\kwd{Next Generation Sequencing}
\kwd{High-Performance Computing}
\end{keyword}

\end{abstractbox}

\end{frontmatter}

\section{Background}
\label{sec:intro}

In the last few years the traditional High Performance Computing (HPC) centres,
such as SciNet at the University of Toronto \cite{scinet-gpc}, have been
witnessing the emergence of increasing amounts of work-flows from non-typical
disciplines in the field of computational science
\cite{SC-HPC-DS_higherEd}. Among those, disciplines related to bioinformatics
appear to be the most prominent in terms of demanding resources and tackling
complex problems related to Next Generation Sequencing (NGS). 
The human genome project \cite{Venter2001},
the human microbiome project \cite{NIH2009},
RNA-Seq to analyze gene expression \cite{Wang2009, Ng2009}
and ChIP-Seq to assess global DNA-binding sites \cite{Wang2009, Ng2009}
are all examples of projects where HPC can be applied.

The advantage of NGS for researchers is  
high-throughput sequencing analysis allowing millions 
of DNA molecules to be read at the same time 
\cite{Schuster2008, Loman2012Corrigendum, Johnson2007}. 
The output of NGS is substantial and it 
can be overwhelming for analyses \cite{Park2009,Mardis2008}.
The difficulties during the analyses can be increased if 
the genome to be studied is more specific. 
We address this problem by combining and utilizing open-source tools 
such as BWA \cite{doi:10.1093/bioinformatics/btp324}, 
SAMtools \cite{doi:10.1093/bioinformatics/btp352}, 
Linux shell and R scripts \cite{citeR} and
techniques commonly employed in the HPC fields.

Chromatin immunoprecipitation coupled to NGS (ChIP-Seq, Figure~\ref{fig:ChIP})
can be utilized to infer function(s) of a protein of interest based on
its chromatin-occupancy profile.
For instance, if the ChIP signal
accumulates closer to a 5' region 
of a gene, we could infer that protein's function may be related to
transcription initiation.
On the other hand,
accumulation of ChIP peaks at the 3' ends would suggest a role in transcription termination. 
However, this has to be tested further since
gene expression can also be coordinated by elements that are not in 
close proximity to the specific gene \cite{CremerCremer2001}. 
Utilization of mock/control samples (untagged) is important 
to account for nonspecific DNA interactions 
with the chromatography resin.

In multi-cellular eukaryotes, the mechanisms of chromatin remodeling involving the
precise delineation of how transcription starts, progresses and ends are poorly
understood. Chromatin remodeling research aids in our understanding of
how these mechanisms affect cellular function in the progression
of disease. ChIP-Seq
is generally used as an initial step to examine the possible role(s) of a protein in chromatin re-modeling \cite{Johnson2007,Schmidt2010}.

Many algorithms and visualization tools are
already set for other model organisms such as human or yeast;
however, none of these applications cater to contig based genomes 
 with only minimal annotations of genes, as is the case in 
\textit{T.thermophila}. These applications also do not directly address 
whether the accumulation of the protein of
interest is in a genic or intergenic region.
Using available tools, such as MACS2 \cite{Zhang2008}, the user
receives the
 peaks coordinates and has to group 
afterwards to assess
the local enrichment within genic and/or intergenic regions. 
Our pipeline, Figure~\ref{fig:Parallel}, tackles this problem 
by offering an one step solution that utilizes the mentioned open sourced tools, 
the available contigs and 
gene annotations files of a genome, providing the user with two 
tables with all the found read accumulations.
The first table contains the reads accumulation in the genic region, 
regions corresponding to the annotated genes. 
The second table is for the accumulation of reads in the intergenic regions. 
The intergenic regions are newly generated each time to account for modifications 
or improvements in the gene annotations.
The obtained results are normalized to the number of clusters that passed Illumina's
``Chastity filter'' also called PF clusters. 
These numbers represent the reads obtained per sample. The normalized 
values are further filtered by using the data obtained from the mock samples.

We present in this paper our computational pipeline to \textbf{R}apidly \textbf{A}nalyse 
\textbf{C}hIP-\textbf{S}eq data for contig based genomes (RACS). 
RACS, which can be run  either in a typical workstation or taking full advantage of
HPC resources, such as, multicore architectures and use of RAMdisk, 
to improve the analysis times making it more efficient,
(see details on Sec.\ref{sec:RACS-HPC}).
This pipeline was developed 
to answer whether the protein of interest localized to a 
gene or an intergenic region.  
RACS was designed in a user friendly manner to accommodate researchers 
with basic knowledge 
in Linux shell and R \cite{citeR}. RACS provides accessible downstream 
analyses of ChIP-Seq data obtained from Illumina instruments.
RACS follows a unique approach to tackle this problem, and is widely applicable
and useful enough to analyse
ChIP-Seq related data from a variety of different organisms generated by NGS.

\section{RACS Pipeline Implementation}

The RACS pipeline is an open source set of shell and R scripts, 
which are organized
in three main categories:
\begin{itemize}
        \item the \textit{core pipeline} tools, which allow the user to
        compute reads differentiating between
        genic and intergenic regions automatically
        \item auxiliary \textit{post-processing} scripts\footnote{Alternatively, 
        we have also included an \textit{auxiliary spreadsheet} that could 
        be used instead of the script to perform the post-processing and normalization 
        manually, as well as, to check against negative controls.}
        for normalization using the "Pass Filter" values
        \item and \textit{utilities} to validate results by visualizing the 
        reads accumulation and run comparisons with other software tools, 
        such as IGV and MACS respectively.
        In particular, the visual representation tools are still under development,
        and will be available in the repository as soon as they are ready.
\end{itemize}

The RACS pipeline will run in any standard workstation with 
a Linux-type operating system.
The pipeline requires the fastq files; the assembly or genomic data, such as, 
    \texttt{T\_thermophila\_June2014.assembly.fasta};
the coordinates for the genic regions, such as, 
    \texttt{T\_thermophila\_June2014.gff3}
(both files can be found at
    \url{http://ciliate.org/index.php/home/downloads}{ciliate.org} \cite{Stover2006});
and the following open source tools:

\begin{itemize}
    \item Burrows-Wheeler Alignment (BWA) version 0.7.13 \cite{doi:10.1093/bioinformatics/btp324}
    \item Sequence Alignment/Map (SAMtools) version 1.3.1 \cite{doi:10.1093/bioinformatics/btp352}
    \item the R statistical language \cite{citeR}
\end{itemize}

Our pipeline is open source, and the scripts are available to download and accessibles
from any of the following repositories\footnote{Both repositories are synchronized, so that the latest version of RACS is available and can be obtained from both of them.}:

    \url{https://gitrepos.scinet.utoronto.ca/public/?a=summary&p=RACS}
	
or

	\url{https://bitbucket.org/mjponce/racs}

The repository includes the core or main scripts placed in the ``core'' directory.
The comparison and auxilary tools are placed in a ``tools'' directory.
And we have also included examples of submission scripts in the ``hpc'' directory,
with PBS \cite{Feng:2007:PUP:1269899.1254906,Steine:2009:PBS:1674636.1674866}
and SLURM \cite{McLay:2011:BPD:2063348.2063360,10.1007/10968987_3}
examples of submission scripts, so that users can take advantage of HPC resources.
Additionally, we have also included a ``datasets'' directory containing examples of data.
Details about the pipeline implementation and how to use it are included in the
`README' file available within the RACS repository.
A generic top-down overview of the pipeline implementation for the data analysis,
is shown in Figure~\ref{fig:CorePipeline}.

Our core scripts do not require any additional R packages; however, the comparison tools, depending on what
format the data to compare with is given, might use some spreadsheet reader package. For instance, we have included
one named \texttt{xlsx} which allows to read propietary formats.
The results of the genic and intergenic regions are generated in two CSV 
(Comma Separated Value) files.
These are standard text ASCII files, which can be read with any typical 
spreadsheet software or R.

We implemented RACS using two data sets of ChIP-Seq data obtained from
the organims \textit{T. thermophila}. The first data set used was 
using the protein Med31-FZZ which is part of the putative Mediator 
complex. The Med31-FZZ data was submitted to a scientific journal and it is under review.
The second data set used is from one of our recent studies \cite{Saettone2018}. Here,
the protein used is Ibd1-FZZ which it was found to be part of multiple chromatin
remodeling complexes.

\subsection{Core Pipeline Tools}

\subsubsection{Determination of the Genic Regions}
\label{sec:DataAnalysis}

To count the amount of reads in each genic region or open reading frame (ORF)
 the core pipeline script was implemented 
using Linux shell commands combined with the usage of
\texttt{BWA} and \texttt{SAMtools}. The input files are the genome of reference 
(\texttt{T\_thermophila\_June2014.assembly.fasta}), 
the gene annotation file (\texttt{T\_thermophila\_June2014.gff3}) and 
the INPUT and IP files obtained from NGS. After the INPUT and IP sequences are 
aligned with the genome and sorted, the script uses a loop to count 
the reads in each ORF and deposits the obtained data in a file named
``\texttt{FINAL.table.}\textit{INPUTfile}-\textit{IPfile}'';
where \textit{INPUTfile} and \textit{IPfile} are the INPUT and IP files respectively. 
Figure~\ref{fig:CorePipeline} depicts a flowchart 
representing the required steps to obtain the final table containing the the number of 
reads found in each of the 
 ORF. 
Details of the processing stages are shown in 
Figure~\ref{fig:Parallel},
in relation to \textit{Tetrahymena} scaffold database and the breakdown of each these steps.

The pipeline is implemented to specifically target the data from the \textit{T.Thermophila}
organism within the \textit{.gff3} file,
if the user would like to consider a different organism, some changes should be done
in the \texttt{table.sh} script located in the core subdirectory.
In particular, the variables \texttt{filter1} and \texttt{filter2}

\begin{verbatim}
    filter1=gene
    filter2="Name=TTHERM_"
\end{verbatim}

should be adjusted correspondingly to the organism of interest.

\subsubsection{Determination of the Intergenic Regions}
\label{sec:intergenicRegions}

The intergenic regions were not available neither determined 
by the standard packages. For this reason, we developed an R script 
to determine these regions.
In this pipeline these sequences are calculated during each run to account 
for further genome actualizations. 
The input for this script are the files generated by 
the genic regions pipeline discussed in the previous section
(ie. ``\texttt{FINAL.table.}\textit{INPUTfile}-\textit{IPfile}'',
the INPUT and IP \textit{.bam} files --which are generated as
intermediate files of the ORF pipeline--) 
plus the gene annotation file (eg. \texttt{T\_thermophila\_June2014.gff3}).
First, the script determines the intergenic regions by 
calculating the beginning and end of each annotated 
gene within each available scaffold and subtracts these values.
The algorithm only reports intergenic regions that are equal or grater to zero.
In the earlier version of the pipeline this constraint was not included,
and in some cases it could result in the pipeline reporting regions 
with negative sizes. We noticed that 92 of these cases were 
presented in our previous study \cite{Saettone2018};
however, we should emphasize that there were not reads present in these regions thus
did not affect these results.
Second, the script uses the newly generated 
intergenic regions to count the
number of reads in each of them. Finally, the data 
is deposited in the intergenic table for each of the
intergenic regions Figure~\ref{fig:CorePipeline} .

\subsection{Post-processing}

\subsubsection{Normalization of Reads accumulation and Enrichment Calculation}
\label{sec:Fold_Enrichment}

To account for differences in the amount of PF clusters (reads) 
presented among samples, the obtained INPUT and IP values were 
normalized by dividing them by the corresponding 
value of the Flowcell summary obtained from the Illumina run or from the 
Total Sequences of the fastQC file. These calculations can be done by 
the script ``\texttt{normalizedORF.sh}'' located in the core directory of RACS.
Alterantively it can also be calculated employing the following two
spreadsheets:
	for the genic regions (\textit{PostProcessing\_Genic.xlsx}), and
	for the intergenic regions (\textit{PostProcessing\_Intergenic.xlsx});
that can be found in the ``datasets'' subdirectory within the RACS repository.

These spreadsheets contain the reads found in the untagged (or mock purification/negative control)
samples in the \textit{Untagged} tab. 
The user can also add the Flowcell summary details in
the \textit{Add\_FCS\_for\_(SAMPLE\_ID)} tab.
The user can manually introduce the read values for the samples being 
analysed in the \textit{Add\_(SAMPLE\_ID)\_ChiP\_Seq} tab. 
In this tab the user can divide the 
number found by RACS by the corresponding PF cluster number found in the previous tab.
This data can be deposited in the ``Normalized\_INPUT or \_IP (FCS)'' columns.
After the required reads normalization the accumulation can be obtained
as the number of IP reads divided by the number of INPUT reads (IP/INPUT). 
This can be deposited in the ``Enrichment\_(N\_IP(FCS)/N\_INPUT(FCS)'' column of the 
same tabs. The obtained values are filtered (\textit{Filter 1}) by the user 
by subtracting the corresponding number  
found in the \textit{Untagged} tab and deposit the 
values in the \textit{Enrichment\_(Minus\_AVERAGE\_untagged} column. 
If there are more than two samples the values can be averaged and values 
that are less than 1.5 can be filtered (\textit{Filter 2}) and deposited in the 
``Enrichment\_Average\_Sample'' tab.
For the ORF table, in this tab there is a column containing the 
Expression profile obtained from the 
\textit{RNA\_Seq} tab. We recommend to copy the filtered cells 
to the \textit{Results} tab. The distribution of the protein of interest 
can be calculated in this tab. 
For the Intergenic table there is a \textit{ORF\_vs\_IGR} (Intergenic) tab where the 
number of regions and reads can be calculated. The number of regions is 
represented by the number of ORF and Intergenic regions that passed the 2 filters.
The number of reads found in the ORF and Intergenic regions
can be calculated by adding all the available values from the 
``Normalized IP (FCS)'' columns and deposit them in the \textit{ORF\_vs\_IGR} tab
of the Intergenic table.

\subsection{Utilities: Validation and Quality Checks}
\label{sec:Qchecks}

To account for biological variability in the wet lab, 
we perform ChIP-Seq using 2 samples for each tagged strain and average their Enrichment. 
To validate the findings, it is important to determine the genic and 
intergenic regions of the untagged (negative control) 
INPUT and IP samples. After this determination, we subtracted the 
obtained average enrichment from 
untagged to the obtained tagged average of the samples. 
Then we filtered for values that had an enrichment greater than or equal 
to 1.5 in the final enrichment column. These are 
the enriched regions and represent where the tagged protein interacts.

\subsubsection{Visualization of Reads accumulation}
The browser IGV \cite{Robinson2011-IGV} can be used to visually inspect and validate the
obtained reads based on their ranked enrichment. The files needed are illustrated 
in Figure~\ref{fig:CorePipeline} and the `README' file included in the RACS' repository. 
MACS2, a main-stream application to call peaks, can 
also be used as specified in \cite{Zhang2008}.
MACS2 uses the same intermediate files (\textit{.bam}) 
obtained from the RACS pipeline, hence it can
be a good reference to be considered for comparison purposes.

\subsection{RACS Performance}
\label{sec:RACS-HPC}
RACS can be run in any normal Linux workstation, however it can also take
advantages of cluster-type environments.
In particular, several stages of RACS can be run using mutlicore architectures
with several threads in parallel.
In addition to that, RAMdisk can be used to speed up file I/O operations, we
have added an optional argument where the user can specify to use RAMdisk if wished.
When we originally developed our pipeline we tested it in our previous
HPC cluster, GPC \cite{scinet-gpc}
consisting of 2.53 GHz Intel Xeon E5540, with 16GB RAM per node (2GB per core).
By comparing the performance of RACS with a typical
workstation we noticed a speed-up factor
among 8 to 12$\times$.
We have also run our pipeline in our newest cluster, Niagara \cite{scinet-niagara},
of 1500 Lenovo SD350 servers each with 40 Intel "Skylake" cores at 2.4 GHz.
Each node of the cluster has 188 GiB / 202 GB RAM per node, for which
we have obtained a speed up of 5 to 10$\times$.
In other words, the whole processing of ORF and IGR for a typical INPUT/IP sample,
took between 1 and 2 hours. In addition to that, in our new system is possible to bundle
40 (80 using multithreading) processes together.

Moreover, this first release of RACS utilizes the basic SAMtools and BAM codes,
however it has been reported that improvements in
processing SAM files could be achieved
using SAMBAMBA \cite{sambamba}.
One of the many advantages of dealing with an open source, modular pipeline like this,
is that it allows interested users to explore this possibility as well, just by
modifying the tool to process SAM files and selecting the one desired.

\section{Results}
\label{sec:results}

\subsection{Case study} 
 
In this section, we will describe how RACS was used to re-analyse
and generate the data
presented in \cite{Saettone2018}.
The model organism used in \cite{Saettone2018} is the protist 
Alveolate \textit{T. thermophila} which is the most experimentally 
amenable member of this taxonomical group.
\textit{T. thermophila} can be used in some cases
to understand the basic biology of the parasitic and disease-causing members of the
Alveolates. Members of \textit{Plasmodium} species that causes malaria 
 \cite{PETERSON2002119, Linder4222,ROELOFS2003132} and 
 other related species that affect
ecosystems \cite{JEU:JEU305} and aquiculture \cite{BUCHMANN2001105} 
can be examined by analogy through our selected model organism. 
In addition, \textit{T. thermophila} has genes that present
homology to human genes \cite{Eisen2006,10.1093/molbev/msz039} and characteristics that makes it
an excellent candidate to study chromatin mainly because of the segregation of
transcriptionally active
and silent chromatin into two distinct nuclei,
macronucleus (MAC) and micronucleus (MIC) respectively
\cite{MARTINDALE1982227}.  
In our recent study we found a protein, Ibd1, that interacts with many chromatin remodeling complexes 
\cite{Saettone2018}.
The \textit{T.thermophila}'s genome \cite{Stover2006} is contig based and contains almost 27 
thousand annotated genes or genic regions \cite{Xiong2012}.
To further the understanding of Ibd1, and 
to contribute to current understanding of how chromatin 
remodeling works, we analysed its localization within the genome by ChIP-Seq 
 (Figure~\ref{fig:ChIP})
\cite{10.1371/journal.pcbi.1003326,citeulike:11192426,Cormier2016,Qin2016}.
This allowed us to understand what genes are being
regulated by Ibd1.

\subsubsection{Pre-Processing of the fastq files and quality assessment}
\label{sec:fastq}
The ChIP samples were processed as described in \cite{Saettone2018}
to make the library preparation using the
TruSeq ChIP-Seq kit (Illumina). 
For the untagged (this study) and Ibd1 \cite{Saettone2018} strains,
libraries were sequenced using the v4 
chemistry in a HiSeq2500 instrument (Illumina) set for High Output mode. The obtained 
read lengths were of 66 base pairs, 6 base pairs corresponded to 
the adapters for demultiplexing. These files were demultiplexed 
in fastq format and the adapters were 
trimmed using the bcl2fastq2 Conversion Software v2.20.0. 
The obtained fastq files for the INPUTS and IP samples were 
assessed by fastQC version 0.11.5 \cite{andrews2010fastqc}.

Each dataset obtained from the ChIP-Seq experiments has 
a sequence of whole cell DNA (INPUT) and DNA sequenced from an immunoprecipitated (IP) sample. 
The Ibd1 NGS data generated in \cite{Saettone2018}
can be found at the following 
Gene Expression Omnibus (GEO) link: \url{https://www.ncbi.nlm.nih.gov/geo/query/acc.cgi?acc=GSE103318}{GSE103318} \cite{Saettone2018}. In addition, untagged
 \textit{T.thermophila} fastq files were generated and they can be found at 
the following GEO link: \url{https://www.ncbi.nlm.nih.gov/geo/query/acc.cgi?acc=GSE125576}{GSE125576} (please use the following token: shqnagaefhilrsz)

We recommend to assess the quality of the data obtained from the NGS. This step is
important to have general information regarding the run.
This fastQC report also helps to understand the alerts 
present in each sample, these alerts do not
necessarily mean that the NGS run failed \cite{andrews2010fastqc}. 
In other words, this step is to verify
if the fastq data has any alerts that have to be addressed
before proceeding to the processing. For example in our case our data for the
\textit{Per base sequence quality} fell into the very good quality reads 
(green) area of the y axis allowing us to avoid quality trimming. 
On the other hand, we obtained a flag for \textit{Overrepresented sequences},
in particular the one that called our attention was 
the sequence containing only the nucleotide N. Since our reads 
are 35-58 base paires long, the allowed maximun mismatch to the genome will be
up to 3 base paires according to the BWA algorithm 
\cite{doi:10.1093/bioinformatics/btp324} 
hence it will not consider these sequences for the alignment.

\subsubsection{Visualization and list of reads}

After, the determination of enriched regions we can further analyse them using
a visualization tool, such as IGV. 
The region of interest can be copied from either the ORF or intergenic table.
This localization corresponds to where the 
protein of interest is localizing with respect to annotated genes 
(see Figure~\ref{fig:genic_vs_MACS2}) or an intergenic region (see Figure~\ref{fig:intergenic_vs_MACS2}).

It is important to note that for Ibd1's ChIP-Seq \cite{Saettone2018} 
data we also used MACS2 a main-stream application to call peaks \cite{Zhang2008}.
The visualization option for MACS2 and RACS are similar in that both provide 
an specific file that can be used for this purpose.
In the case of RACS, our pipeline uses \textit{.bam} and \textit{.fai} files which are
generated within the ORF part of the pipeline (see Figure~\ref{fig:CorePipeline}).
Such \textit{.bam} files can be opened in IGV, although the \textit{.fai} (index) file will not,
however both files should be present in the same directory.
The required files for IGV visualization are depicted in Figure~\ref{fig:CorePipeline}.
In addition, the \textit{.bam} files generated by RACS can be used as input for MACS2.  
When compared the MACS2 visualization file to the RACS \textit{.bam} files using IGV 
(Figure~\ref{fig:genic_vs_MACS2}), we observed that the RACS files provide 
a visual of the portion enriched. Here we observed that the IP samples are
clearly enriched regions showing peaks when compared to the INPUT samples. 
This can be determined by noting the numbers shown in each of the IGV tracks,
which represent its corresponding reads accumulation.

The output lists provided by RACS are segregated into two \textit{.csv} files.
The first file contain the ORF and the second the intergenic regions. Both lists 
contain the all reads obtained from the INPUT and IP samples. 
This obtained data should be filtered with the data obtained from Mock samples.
We found that the output of MACS2 provides a list of peaks. 
MACS2 does not classify the peaks based on the localization 
to a gene or an intergenic region as RACS does. However, this can be addressed using 
BEDTools \cite{10.1093/bioinformatics/btq033} after MACS2 analysis.
The datasets generated by MACS2 and RACS can be found at 
\url{https://www.ncbi.nlm.nih.gov/geo/query/acc.cgi?acc=GSE103318}{GSE103318}.

\subsubsection{Mock samples facilitate the analysis}
In \cite{Saettone2018} we found that the majority of genes regulated by 
Ibd1 were highly expressed housekeeping genes.
For that analysis 
untagged samples were not employed, instead a cut-off based on 
accumulation of reads was implemented.
However, since the 
cut-off was arbitrary there was some degree of uncertainty in regards of 
its astringency.
To overcome this limitation and further facilitate the analysis,
for this study 
we performed ChIP-Seq using untagged control \textit{T.thermophila} cells.
The addition of two mock ChIP-Seq 
replicas for this study (untagged \textit{T.thermophila}) 
enhanced RACS' ability to discriminate non specific DNA binding.
In addition, the use of mock ChIP-Seq samples eliminated the uncertainty 
associated with using the arbitrary cut-off.
This newly generated analysis will be used as a model for this case study.
Between both the analysis presented in \cite{Saettone2018}, and this new 
analysis (RACS),
there are not major statistical differences regarding Ibd1 localization to
genes that are highly, moderate, or low to no-expressed.
The statistical analysis presented in Figure~\ref{fig:Table} 
shows that the hypothesis generated 
in \cite{Saettone2018} regarding an
Ibd1 function related to transcription of highly expressed genes stands. 
The calculation of this result can be found in the 
Result tab of the ORF table (\IBDorfURL).

\subsubsection{RACS aids in the determination of the protein of interest function}

By segregating the protein's localization between genic and intergenic, the 
determination of its function is rapid.
After analizing the Ibd1's data with RACS, tables with the total number of
reads found in each of the 26,996 ORF and 27,780 intergenic regions were generated \cite{Saettone2018}.

From the ORF and intergenic tables we observed that Ibd1 localizes to 
more individual intergenic regions than genic 
regions (Figure~\ref{fig:Ibd1_distribution}-A).
However, the majority of reads accumulation are in the genic regions
(Figure~\ref{fig:Ibd1_distribution}-B),
suggesting that Ibd1 localization is within the ORF and its function maybe
related to transcription 
regulation.

The function of Ibd1 was further inferred based on its gene-specific 
localization. From the ORF table (\IBDorfURL)
we observed that Ibd1 mostly localizes to 
genes that are highly expressed and related to housekeeping function; such as, 
cellular function, translation, gene expression, biogenesis, 
cytoplasmic translation among others (see Figure~\ref{fig:Ibd1_GO}). 
The functions are based on GO annotations for biological 
process \cite{ashburner2000gene}. The calculation of this result can be found in the 
\textit{Gene\_Ontology} tab of the ORF table (\IBDorfURL).

\subsubsection{Outliers}

We found a great accumulation of the following three ORFs: TTHERM\_02141639, 
TTHERM\_02641280, TTHERM\_02653301; in the tagged and untagged ChIP-Seq samples.
Since we found this DNA accumulated in the untagged and tagged samples we concluded that
this event may be due to non-specific binding to the agarose matrix. 
For Ibd1, the first 2 were filter out after the subtraction step described in the
``Utilities: Validation and Quality Checks'' section (Sec.~\ref{sec:Qchecks}),
and the third region seems to be controled by Ibd1.

\subsection{Performance}
By implementing this pipeline as described here, we obtained roughly a factor of
$4\times$ faster in comparison to a serial and non-I/O optimized (ie. not using RAMdisk),
in a equivalent hardware to the node used in the cluster.
This is something we have also observed by using similar techniques (eg. RAMdisk) in
other type of bio-informatics pipelines where the hierarchy of the computational scales
is dominated by the I/O parts of the code.
Moreover, we processed a second set of data, that was roughly 3 times larger than
the original data --which would not fit in memory ($>64$GB)--,
utilizing a more modern node (i7 core) with a solid-state device (SSD),
we were able to furher reduce the processing time approximately by another factor of $\sim 4$.
This type of trend is typical in cases where performance is dominated by computations and I/O operations (eg. reading and writting files),
for which the combination of faster processing plus faster access to the data is essential for improving the overall performance.
Nevertheless, we should emphasize that even when RAMdisk or a SSD can be a solution that could in principle be
thrown to similar type of problems, ie. intensively I/O demanding ones,
the best approach would always be to try to mitigate and reduce as much as possible the
I/O operations, as these usually represent the slowest part in any computational implementation.

\section{Discussion}
\label{discussion}

In this paper, we have presented a pipeline implementation
utilizing open source tools. This pipeline aids in the rapid analysis of ChIP-Seq data
grouping the reads accumulation from gene or intergenic regions throughout
a contig based genome. The objective is to infer protein function based on
its chromatin occupancy.
This pipeline has been applied to \textit{Tetrahymena thermophila}'s
ChIP-Seq data, but its application can be extended to
ChIP-Seq datasets generated in any other organisms.

We called peaks using MACS2, the output of this application 
was efficient; however, did not elucidate whether the protein 
of interest localizes to a genic or an intergenic region. 
Different to MACS2 that calls all peaks regardless of their 
position in a genic or intergenic region, RACS was designed with that
very same goal in mind, hence the diferentiation in the modules that
are offered in our pipeline.
We should also notice that other tools such as BEDTools,
can be used to perform this task in combination with MACS2.

Without the need of any additional ``external'' software, RACS calls reads 
and segregates them on a genic or intergenic regions table.
Thus our pipeline is more suitable to answer biological questions
regarding function. 
We believe that MACS2 is complementary to RACS, in that 
both are peaks and reads accumulation predictors.
As shown in Figure~\ref{fig:genic_vs_MACS2}, by obtaining a top hit from RACS this 
regions can be compared or assessed by MACS2 at the same time to 
obtained a more robust result.

It is important noting that some regions presented in the processed table 
have very few reads after substracting the values obtained from untagged samples   .
For instance, a sample that has 2 reads in the INPUT and 10 reads in the IP 
will return an enrichment of 5 and it may pass the filter of 1.5X enrichment. 
Even when there are many computational tools available for processing 
ChIP-Seq data, RACS is particularly suitable for contig genomes with limited annotations.
Recently several ChIP-Seq studies \cite{wang2017n6,kataoka2015phosphorylation} 
have emerged for \textit{T.Thermophila}.
However, there is a lack of standardized computational methods for 
this model organism, hence it becomes difficult to reliably reach at 
the same conclusions when replicating the findings.
Our tool is the first effort in \textit{T.Thermophila} 
to provide a community resource for genome-wide ChIP-Seq studies, 
therefore it will contribute in standardizing the analyses.
Other tools, such as, MACS2 and metagene using BEDTools analysis \cite{10.1093/bioinformatics/btq033}
complement RACS.

Last but not least, we must also note that our tool is of a broad 
utility and can be easily used for any other model system for ChIP-Seq analyses.

\section{Conclusions}

\label{sec:conclu}

RACS is an excellent tool for genomes that are contig-based and/or have poor annotations,
it permits the segregation of reads accumulation after ChIP-Seq processing.
RACS is complementary to other tools, such as MACS2,
as it can help to discriminate complex regions improving the overall analysis.

RACS offers an alternative tool with a different approach focused on a
simple, modular and open approach. RACS offers a 
versatile, agile and modular pipeline that cover many of the steps needed in 
the process of analysing ChIP-Seq data.

The pipeline uses HPC tools, such as RAMdisk or batch processing via scheduling
in cluster type environments, so that the data analysis can be done for 
large datasets.
The scripts are reusable and generic enough that can be simply modified and
utilized in other pipelines as well.

The modular approach we followed when developing RACS, also allows for
future developments as this pipeline could be easily ported as a backend
of a web interface, or a \textit{gateway} portal, serving a larger group
of researchers from different disciplines.

\section*{Declarations}

\subsection*{\textbf{Ethics approval and consent to participate}}
Not applicable.

\subsection*{\textbf{Consent for publication}}
Not applicable.

\subsection*{\textbf{Availability of data and materials}}
ChIP-Seq data used to develop this methodolgy can be found online at Gene
Expression Omnibus (GEO) \url{http://www.ncbi.nlm.nih.gov/geo/} GSE103318 and GSE125576.
\\
NGS and peak files produced in this study were deposited at
https://www.ncbi.nlm.nih.gov/geo/ with unique identifiers GSE103318 and GSE125576 (please use the following token: shqnagaefhilrsz).
\\
Direct links:
	\url{http://www.ncbi.nlm.nih.gov/geo/query/acc.cgi?acc=GSE103318}
and
	\url{http://www.ncbi.nlm.nih.gov/geo/query/acc.cgi?acc=GSE125576}.
\\
RACS pipeline (including a README file with instructions in how to use the scripts)
can be accessed through any of the following repositories:

       \url{https://gitrepos.scinet.utoronto.ca/public/?a=summary&p=RACS}

       \url{https://bitbucket.org/mjponce/racs}.

\subsection*{\textbf{Competing interests}}
  The authors declare that they have no competing interests.

\subsection*{\textbf{Funding}}
Work in the Fillingham laboratory was supported by Natural Sciences and Engineering
Research Council of Canada (NSERC) Discovery Grants RGPIN-2015-06448.
SciNet is funded by: the Canada Foundation for Innovation under the auspices of
Compute Canada; the Government of Ontario; Ontario Research Fund—Research
Excellence; and the University of Toronto.

\subsection*{\textbf{Author's contributions}}
AS defined the alignment steps using BWA and SAMtools and the reads
extraction methodology and MP automated these steps by developing the scripts.
AS conceived the reads extraction methodology for the intergenic regions
and MP developed and automated it.
AS and MP analysed the data, designed the study and wrote the manuscript.
AS and SNS prepared the ChIP samples from untagged strains.
JF conceived the study, coordinated and edited the manuscript.
All authors read and approved the final
manuscript.

\subsection*{\textbf{Acknowledgements}}
We acknowledge Tanja Durbic from the Donnelly Sequencing
Centre at the University of Toronto for sequencing our ChIP samples.

\begin{backmatter}

\bibliographystyle{bmc-mathphys} \bibliography{references}


\begin{thebibliography}{44}
\ifx \bisbn   \undefined \def \bisbn  #1{ISBN #1}\fi
\ifx \binits  \undefined \def \binits#1{#1}\fi
\ifx \bauthor  \undefined \def \bauthor#1{#1}\fi
\ifx \batitle  \undefined \def \batitle#1{#1}\fi
\ifx \bjtitle  \undefined \def \bjtitle#1{#1}\fi
\ifx \bvolume  \undefined \def \bvolume#1{\textbf{#1}}\fi
\ifx \byear  \undefined \def \byear#1{#1}\fi
\ifx \bissue  \undefined \def \bissue#1{#1}\fi
\ifx \bfpage  \undefined \def \bfpage#1{#1}\fi
\ifx \blpage  \undefined \def \blpage #1{#1}\fi
\ifx \burl  \undefined \def \burl#1{\textsf{#1}}\fi
\ifx \doiurl  \undefined \def \doiurl#1{\textsf{#1}}\fi
\ifx \betal  \undefined \def \betal{\textit{et al.}}\fi
\ifx \binstitute  \undefined \def \binstitute#1{#1}\fi
\ifx \binstitutionaled  \undefined \def \binstitutionaled#1{#1}\fi
\ifx \bctitle  \undefined \def \bctitle#1{#1}\fi
\ifx \beditor  \undefined \def \beditor#1{#1}\fi
\ifx \bpublisher  \undefined \def \bpublisher#1{#1}\fi
\ifx \bbtitle  \undefined \def \bbtitle#1{#1}\fi
\ifx \bedition  \undefined \def \bedition#1{#1}\fi
\ifx \bseriesno  \undefined \def \bseriesno#1{#1}\fi
\ifx \blocation  \undefined \def \blocation#1{#1}\fi
\ifx \bsertitle  \undefined \def \bsertitle#1{#1}\fi
\ifx \bsnm \undefined \def \bsnm#1{#1}\fi
\ifx \bsuffix \undefined \def \bsuffix#1{#1}\fi
\ifx \bparticle \undefined \def \bparticle#1{#1}\fi
\ifx \barticle \undefined \def \barticle#1{#1}\fi
\ifx \bconfdate \undefined \def \bconfdate #1{#1}\fi
\ifx \botherref \undefined \def \botherref #1{#1}\fi
\ifx \url \undefined \def \url#1{\textsf{#1}}\fi
\ifx \bchapter \undefined \def \bchapter#1{#1}\fi
\ifx \bbook \undefined \def \bbook#1{#1}\fi
\ifx \bcomment \undefined \def \bcomment#1{#1}\fi
\ifx \oauthor \undefined \def \oauthor#1{#1}\fi
\ifx \citeauthoryear \undefined \def \citeauthoryear#1{#1}\fi
\ifx \endbibitem  \undefined \def \endbibitem {}\fi
\ifx \bconflocation  \undefined \def \bconflocation#1{#1}\fi
\ifx \arxivurl  \undefined \def \arxivurl#1{\textsf{#1}}\fi
\csname PreBibitemsHook\endcsname

%%% 1
\bibitem{scinet-gpc}
\begin{barticle}
\bauthor{\bsnm{Loken}, \binits{C.}},
\bauthor{\bsnm{Gruner}, \binits{D.}},
\bauthor{\bsnm{Groer}, \binits{L.}},
\bauthor{\bsnm{Peltier}, \binits{R.}},
\bauthor{\bsnm{Bunn}, \binits{N.}},
\bauthor{\bsnm{Craig}, \binits{M.}},
\bauthor{\bsnm{Henriques}, \binits{T.}},
\bauthor{\bsnm{Dempsey}, \binits{J.}},
\bauthor{\bsnm{Yu}, \binits{C.-H.}},
\bauthor{\bsnm{Chen}, \binits{J.}},
\bauthor{\bsnm{Dursi}, \binits{L.J.}},
\bauthor{\bsnm{Chong}, \binits{J.}},
\bauthor{\bsnm{Northrup}, \binits{S.}},
\bauthor{\bsnm{Pinto}, \binits{J.}},
\bauthor{\bsnm{Knecht}, \binits{N.}},
\bauthor{\bsnm{Zon}, \binits{R.V.}}:
\batitle{{SciNet: Lessons Learned from Building a Power-efficient Top-20 System
  and Data Centre}}.
\bjtitle{Journal of Physics: Conference Series}
\bvolume{256}(\bissue{1}),
\bfpage{012026}
(\byear{2010})
\end{barticle}
\endbibitem

%%% 2
\bibitem{SC-HPC-DS_higherEd}
\begin{barticle}
\bauthor{\bsnm{Ponce}, \binits{M.}},
\bauthor{\bsnm{Spence}, \binits{E.}},
\bauthor{\bsnm{Gruner}, \binits{D.}},
\bauthor{\bparticle{van} \bsnm{Zon}, \binits{R.}}:
\batitle{Scientific computing, high-performance computing and data science in
  higher education}.
\bjtitle{Journal of Computational Science Education}
(\byear{2019}).
doi:\doiurl{10.22369/issn.2153-4136/10/1/5}.
\arxivurl{1604.05676}
\end{barticle}
\endbibitem

%%% 3
\bibitem{Venter2001}
\begin{barticle}
\bauthor{\bsnm{Venter}, \binits{e.a.} \bsuffix{J.~Craig}}:
\batitle{The sequence of the human genome}.
\bjtitle{Science}
\bvolume{291}(\bissue{5507}),
\bfpage{1304}--\blpage{1351}
(\byear{2001}).
doi:\doiurl{10.1126/science.1058040}
\end{barticle}
\endbibitem

%%% 4
\bibitem{NIH2009}
\begin{botherref}
\oauthor{\bsnm{{The NIH HMP Working Group, Jane Peterson, et al}}}:
The nih human microbiome project.
Genome Research
\textbf{19}(12)
(2009).
doi:\doiurl{10.1101/gr.096651.109}
\end{botherref}
\endbibitem

%%% 5
\bibitem{Wang2009}
\begin{botherref}
\oauthor{\bsnm{{Wang, Zhong; Gerstein, Mark; Snyder, Michael}}}:
Rna-seq: a revolutionary tool for transcriptomics.
Genetics
\textbf{10}(1)
(2009).
doi:\doiurl{10.1038/nrg2484}
\end{botherref}
\endbibitem

%%% 6
\bibitem{Ng2009}
\begin{botherref}
\oauthor{\bsnm{{Ng, Sarah B; Turner, Emily H; Robertson, Peggy D; Flygare,
  Steven D; Bigham, Abigail W; et al.}}}:
Targeted capture and massively parallel sequencing of 12 human exomes.
Nature
\textbf{461}(7261)
(2009)
\end{botherref}
\endbibitem

%%% 7
\bibitem{Schuster2008}
\begin{botherref}
\oauthor{\bsnm{Schuster}, \binits{S.C.}}:
Next-generation sequencing transforms today's biology.
Nature Methods
\textbf{5}
(2008).
doi:\doiurl{10.1038/nmeth1156}
\end{botherref}
\endbibitem

%%% 8
\bibitem{Loman2012Corrigendum}
\begin{botherref}
\oauthor{\bsnm{Loman}, \binits{N.}},
\oauthor{\bsnm{Misra}, \binits{R.}},
\oauthor{\bsnm{Dallman}, \binits{T.}},
\oauthor{\bsnm{Constantinidou}, \binits{C.}},
\oauthor{\bsnm{Gharbia}, \binits{S.}},
\oauthor{\bsnm{Wain}, \binits{J.}},
\oauthor{\bsnm{Pallen}, \binits{M.}}:
Corrigendum: Performance comparison of benchtop high-throughput sequencing
  platforms.
Nature Biotechnology
\textbf{30}
(2012).
doi:\doiurl{10.1038/nbt0612-562f}
\end{botherref}
\endbibitem

%%% 9
\bibitem{Johnson2007}
\begin{barticle}
\bauthor{\bsnm{Johnson}, \binits{D.S.}},
\bauthor{\bsnm{Mortazavi}, \binits{A.}},
\bauthor{\bsnm{Myers}, \binits{R.M.}},
\bauthor{\bsnm{Wold}, \binits{B.}}:
\batitle{Genome-wide mapping of in vivo protein-dna interactions}.
\bjtitle{Science}
\bvolume{316}(\bissue{5830}),
\bfpage{1497}--\blpage{1502}
(\byear{2007}).
doi:\doiurl{10.1126/science.1141319}.
\arxivurl{http://science.sciencemag.org/content/316/5830/1497.full.pdf}
\end{barticle}
\endbibitem

%%% 10
\bibitem{Park2009}
\begin{botherref}
\oauthor{\bsnm{Park}, \binits{P.J.}}:
Chip–seq: advantages and challenges of a maturing technology.
Nature Reviews Genetics
\textbf{10}
(2009).
doi:\doiurl{10.1038/nrg2641}
\end{botherref}
\endbibitem

%%% 11
\bibitem{Mardis2008}
\begin{botherref}
\oauthor{\bsnm{Mardis}, \binits{E.R.}}:
The impact of next-generation sequencing technology on genetics.
Trends in Genetics
\textbf{24}
(2008).
doi:\doiurl{0.1016/j.tig.2007.12.007}
\end{botherref}
\endbibitem

%%% 12
\bibitem{doi:10.1093/bioinformatics/btp324}
\begin{barticle}
\bauthor{\bsnm{Li}, \binits{H.}},
\bauthor{\bsnm{Durbin}, \binits{R.}}:
\batitle{{Fast and accurate short read alignment with Burrows–Wheeler
  transform}}.
\bjtitle{Bioinformatics}
\bvolume{25}(\bissue{14}),
\bfpage{1754}
(\byear{2009}).
doi:\doiurl{10.1093/bioinformatics/btp324}
\end{barticle}
\endbibitem

%%% 13
\bibitem{doi:10.1093/bioinformatics/btp352}
\begin{barticle}
\bauthor{\bsnm{Li}, \binits{H.}},
\bauthor{\bsnm{Handsaker}, \binits{B.}},
\bauthor{\bsnm{Wysoker}, \binits{A.}},
\bauthor{\bsnm{Fennell}, \binits{T.}},
\bauthor{\bsnm{Ruan}, \binits{J.}},
\bauthor{\bsnm{Homer}, \binits{N.}},
\bauthor{\bsnm{Marth}, \binits{G.}},
\bauthor{\bsnm{Abecasis}, \binits{G.}},
\bauthor{\bsnm{Durbin}, \binits{R.}}:
\batitle{{The Sequence Alignment/Map format and SAMtools}}.
\bjtitle{Bioinformatics}
\bvolume{25}(\bissue{16}),
\bfpage{2078}
(\byear{2009}).
doi:\doiurl{10.1093/bioinformatics/btp352}
\end{barticle}
\endbibitem

%%% 14
\bibitem{citeR}
\begin{bbook}
\bauthor{\bsnm{{R Core Team}}}:
\bbtitle{R: A Language and Environment for Statistical Computing}.
\bpublisher{R Foundation for Statistical Computing},
\blocation{Vienna, Austria}
(\byear{2016}).
\bcomment{R Foundation for Statistical Computing}.
\burl{https://www.R-project.org/}
\end{bbook}
\endbibitem

%%% 15
\bibitem{CremerCremer2001}
\begin{barticle}
\bauthor{\bsnm{Cremer}, \binits{T.}},
\bauthor{\bsnm{Cremer}, \binits{C.}}:
\batitle{Chromosome territories, nuclear architecture and gene regulation in
  mammalian cells}.
\bjtitle{Nature Reviews.Genetics}
\bvolume{2}(\bissue{4}),
\bfpage{292}--\blpage{301}
(\byear{2001}).
\bcomment{Copyright - Copyright Nature Publishing Group Apr 2001; Last updated
  - 2013-01-27}
\end{barticle}
\endbibitem

%%% 16
\bibitem{Schmidt2010}
\begin{barticle}
\bauthor{\bsnm{Schmidt}, \binits{D.}},
\bauthor{\bsnm{Wilson}, \binits{M.D.}},
\bauthor{\bsnm{Ballester}, \binits{B.}},
\bauthor{\bsnm{Schwalie}, \binits{P.C.}},
\bauthor{\bsnm{Brown}, \binits{G.D.}},
\bauthor{\bsnm{Marshall}, \binits{A.}},
\bauthor{\bsnm{Kutter}, \binits{C.}},
\bauthor{\bsnm{Watt}, \binits{S.}},
\bauthor{\bsnm{Martinez-Jimenez}, \binits{C.P.}},
\bauthor{\bsnm{Mackay}, \binits{S.}},
\bauthor{\bsnm{Talianidis}, \binits{I.}},
\bauthor{\bsnm{Flicek}, \binits{P.}},
\bauthor{\bsnm{Odom}, \binits{D.T.}}:
\batitle{Five-vertebrate chip-seq reveals the evolutionary dynamics of
  transcription factor binding}.
\bjtitle{Science}
\bvolume{328}(\bissue{5981}),
\bfpage{1036}--\blpage{1040}
(\byear{2010}).
doi:\doiurl{10.1126/science.1186176}
\end{barticle}
\endbibitem

%%% 17
\bibitem{Zhang2008}
\begin{barticle}
\bauthor{\bsnm{Zhang}, \binits{Y.}},
\bauthor{\bsnm{Liu}, \binits{T.}},
\bauthor{\bsnm{Meyer}, \binits{C.A.}},
\bauthor{\bsnm{Eeckhoute}, \binits{J.}},
\bauthor{\bsnm{Johnson}, \binits{D.S.}},
\bauthor{\bsnm{Bernstein}, \binits{B.E.}},
\bauthor{\bsnm{Nusbaum}, \binits{C.}},
\bauthor{\bsnm{Myers}, \binits{R.M.}},
\bauthor{\bsnm{Brown}, \binits{M.}},
\bauthor{\bsnm{Li}, \binits{W.}},
\bauthor{\bsnm{Liu}, \binits{X.S.}}:
\batitle{Model-based analysis of chip-seq (macs)}.
\bjtitle{Genome Biology}
\bvolume{9}(\bissue{9}),
\bfpage{137}
(\byear{2008}).
doi:\doiurl{10.1186/gb-2008-9-9-r137}
\end{barticle}
\endbibitem

%%% 18
\bibitem{Stover2006}
\begin{barticle}
\bauthor{\bsnm{Stover}, \binits{N.A.}},
\bauthor{\bsnm{Krieger}, \binits{C.J.}},
\bauthor{\bsnm{Binkley}, \binits{G.}},
\bauthor{\bsnm{Dong}, \binits{Q.}},
\bauthor{\bsnm{Fisk}, \binits{D.G.}},
\bauthor{\bsnm{Nash}, \binits{R.}},
\bauthor{\bsnm{Sethuraman}, \binits{A.}},
\bauthor{\bsnm{Weng}, \binits{S.}},
\bauthor{\bsnm{Cherry}, \binits{J.M.}}:
\batitle{Tetrahymena genome database (tgd): a new genomic resource for
  tetrahymena thermophila research}.
\bjtitle{Nucleic Acids Research}
\bvolume{34}(\bissue{suppl\_1}),
\bfpage{500}--\blpage{503}
(\byear{2006}).
doi:\doiurl{10.1093/nar/gkj054}
\end{barticle}
\endbibitem

%%% 19
\bibitem{Feng:2007:PUP:1269899.1254906}
\begin{barticle}
\bauthor{\bsnm{Feng}, \binits{H.}},
\bauthor{\bsnm{Misra}, \binits{V.}},
\bauthor{\bsnm{Rubenstein}, \binits{D.}}:
\batitle{Pbs: A unified priority-based scheduler}.
\bjtitle{SIGMETRICS Perform. Eval. Rev.}
\bvolume{35}(\bissue{1}),
\bfpage{203}--\blpage{214}
(\byear{2007}).
doi:\doiurl{10.1145/1269899.1254906}
\end{barticle}
\endbibitem

%%% 20
\bibitem{Steine:2009:PBS:1674636.1674866}
\begin{bchapter}
\bauthor{\bsnm{Steine}, \binits{M.}},
\bauthor{\bsnm{Bekooij}, \binits{M.}},
\bauthor{\bsnm{Wiggers}, \binits{M.}}:
\bctitle{A priority-based budget scheduler with conservative dataflow model}.
In: \bbtitle{Proceedings of the 2009 12th Euromicro Conference on Digital
  System Design, Architectures, Methods and Tools}.
\bsertitle{DSD '09},
pp. \bfpage{37}--\blpage{44}.
\bpublisher{IEEE Computer Society},
\blocation{Washington, DC, USA}
(\byear{2009}).
doi:\doiurl{10.1109/DSD.2009.148}.
\burl{https://doi.org/10.1109/DSD.2009.148}
\end{bchapter}
\endbibitem

%%% 21
\bibitem{McLay:2011:BPD:2063348.2063360}
\begin{bchapter}
\bauthor{\bsnm{McLay}, \binits{R.}},
\bauthor{\bsnm{Schulz}, \binits{K.W.}},
\bauthor{\bsnm{Barth}, \binits{W.L.}},
\bauthor{\bsnm{Minyard}, \binits{T.}}:
\bctitle{Best practices for the deployment and management of production hpc
  clusters}.
In: \bbtitle{State of the Practice Reports}.
\bsertitle{SC '11},
pp. \bfpage{9}--\blpage{1911}.
\bpublisher{ACM},
\blocation{New York, NY, USA}
(\byear{2011}).
doi:\doiurl{10.1145/2063348.2063360}.
\burl{http://doi.acm.org/10.1145/2063348.2063360}
\end{bchapter}
\endbibitem

%%% 22
\bibitem{10.1007/10968987_3}
\begin{bchapter}
\bauthor{\bsnm{Yoo}, \binits{A.B.}},
\bauthor{\bsnm{Jette}, \binits{M.A.}},
\bauthor{\bsnm{Grondona}, \binits{M.}}:
\bctitle{Slurm: Simple linux utility for resource management}.
In: \beditor{\bsnm{Feitelson}, \binits{D.}},
\beditor{\bsnm{Rudolph}, \binits{L.}},
\beditor{\bsnm{Schwiegelshohn}, \binits{U.}} (eds.)
\bbtitle{Job Scheduling Strategies for Parallel Processing},
pp. \bfpage{44}--\blpage{60}.
\bpublisher{Springer},
\blocation{Berlin, Heidelberg}
(\byear{2003})
\end{bchapter}
\endbibitem

%%% 23
\bibitem{Saettone2018}
\begin{barticle}
\bauthor{\bsnm{Saettone}, \binits{A.}},
\bauthor{\bsnm{Garg}, \binits{J.}},
\bauthor{\bsnm{Lambert}, \binits{J.-P.}},
\bauthor{\bsnm{Nabeel-Shah}, \binits{S.}},
\bauthor{\bsnm{Ponce}, \binits{M.}},
\bauthor{\bsnm{Burtch}, \binits{A.}},
\bauthor{\bsnm{Thuppu~Mudalige}, \binits{C.}},
\bauthor{\bsnm{Gingras}, \binits{A.-C.}},
\bauthor{\bsnm{Pearlman}, \binits{R.E.}},
\bauthor{\bsnm{Fillingham}, \binits{J.}}:
\batitle{The bromodomain-containing protein ibd1 links multiple
  chromatin-related protein complexes to highly expressed genes in tetrahymena
  thermophila}.
\bjtitle{Epigenetics {\&} Chromatin}
\bvolume{11}(\bissue{1}),
\bfpage{10}
(\byear{2018}).
doi:\doiurl{10.1186/s13072-018-0180-6}
\end{barticle}
\endbibitem

%%% 24
\bibitem{Robinson2011-IGV}
\begin{barticle}
\bauthor{\bsnm{Robinson}, \binits{J.T.}},
\bauthor{\bsnm{Thorvaldsd{\'o}ttir}, \binits{H.}},
\bauthor{\bsnm{Winckler}, \binits{W.}},
\bauthor{\bsnm{Guttman}, \binits{M.}},
\bauthor{\bsnm{Lander}, \binits{E.S.}},
\bauthor{\bsnm{Getz}, \binits{G.}},
\bauthor{\bsnm{Mesirov}, \binits{J.P.}}:
\batitle{{Integrative Genomics Viewer}}.
\bjtitle{{Nature Biotechnology}}
\bvolume{29}(\bissue{1}),
\bfpage{24}--\blpage{26}
(\byear{2011})
\end{barticle}
\endbibitem

%%% 25
\bibitem{scinet-niagara}
\begin{botherref}
\oauthor{\bsnm{{Ponce et al.}}}:
{Deploying a Top-100 Supercomputer for Large Parallel Workloads: the Niagara
  Supercomputer}.
{in prep.}
(2019)
\end{botherref}
\endbibitem

%%% 26
\bibitem{sambamba}
\begin{barticle}
\bauthor{\bsnm{Tarasov}, \binits{A.}},
\bauthor{\bsnm{Vilella}, \binits{A.J.}},
\bauthor{\bsnm{Cuppen}, \binits{E.}},
\bauthor{\bsnm{Nijman}, \binits{I.J.}},
\bauthor{\bsnm{Prins}, \binits{P.}}:
\batitle{Sambamba: fast processing of ngs alignment formats}.
\bjtitle{Bioinformatics}
\bvolume{31}(\bissue{12}),
\bfpage{2032}--\blpage{2034}
(\byear{2015}).
doi:\doiurl{10.1093/bioinformatics/btv098}
\end{barticle}
\endbibitem

%%% 27
\bibitem{PETERSON2002119}
\begin{barticle}
\bauthor{\bsnm{Peterson}, \binits{D.S.}},
\bauthor{\bsnm{Gao}, \binits{Y.}},
\bauthor{\bsnm{Asokan}, \binits{K.}},
\bauthor{\bsnm{Gaertig}, \binits{J.}}:
\batitle{The circumsporozoite protein of plasmodium falciparum is expressed and
  localized to the cell surface in the free-living ciliate tetrahymena
  thermophila}.
\bjtitle{Molecular and Biochemical Parasitology}
\bvolume{122}(\bissue{2}),
\bfpage{119}--\blpage{126}
(\byear{2002}).
doi:\doiurl{10.1016/S0166-6851(02)00079-8}
\end{barticle}
\endbibitem

%%% 28
\bibitem{Linder4222}
\begin{barticle}
\bauthor{\bsnm{Linder}, \binits{J.U.}},
\bauthor{\bsnm{Engel}, \binits{P.}},
\bauthor{\bsnm{Reimer}, \binits{A.}},
\bauthor{\bsnm{Kr{\"u}ger}, \binits{T.}},
\bauthor{\bsnm{Plattner}, \binits{H.}},
\bauthor{\bsnm{Schultz}, \binits{A.}},
\bauthor{\bsnm{Schultz}, \binits{J.E.}}:
\batitle{Guanylyl cyclases with the topology of mammalian adenylyl cyclases and
  an n-terminal p-type atpase-like domain in paramecium, tetrahymena and
  plasmodium}.
\bjtitle{The EMBO Journal}
\bvolume{18}(\bissue{15}),
\bfpage{4222}--\blpage{4232}
(\byear{1999}).
doi:\doiurl{10.1093/emboj/18.15.4222}.
\arxivurl{http://emboj.embopress.org/content/18/15/4222.full.pdf}
\end{barticle}
\endbibitem

%%% 29
\bibitem{ROELOFS2003132}
\begin{barticle}
\bauthor{\bsnm{Roelofs}, \binits{J.}},
\bauthor{\bsnm{Smith}, \binits{J.L.}},
\bauthor{\bsnm{Haastert}, \binits{P.J.M.V.}}:
\batitle{cgmp signalling: different ways to create a pathway}.
\bjtitle{Trends in Genetics}
\bvolume{19}(\bissue{3}),
\bfpage{132}--\blpage{134}
(\byear{2003}).
doi:\doiurl{10.1016/S0168-9525(02)00044-6}
\end{barticle}
\endbibitem

%%% 30
\bibitem{JEU:JEU305}
\begin{barticle}
\bauthor{\bsnm{Tang}, \binits{Y.Z.}},
\bauthor{\bsnm{Egerton}, \binits{T.A.}},
\bauthor{\bsnm{Kong}, \binits{L.}},
\bauthor{\bsnm{Marshall}, \binits{H.G.}}:
\batitle{Morphological variation and phylogenetic analysis of the
  dinoflagellate gymnodinium aureolum from a tributary of chesapeake bay}.
\bjtitle{Journal of Eukaryotic Microbiology}
\bvolume{55}(\bissue{2}),
\bfpage{91}--\blpage{99}
(\byear{2008}).
doi:\doiurl{10.1111/j.1550-7408.2008.00305.x}
\end{barticle}
\endbibitem

%%% 31
\bibitem{BUCHMANN2001105}
\begin{barticle}
\bauthor{\bsnm{Buchmann}, \binits{K.}},
\bauthor{\bsnm{Sigh}, \binits{J.}},
\bauthor{\bsnm{Nielsen}, \binits{C.V.}},
\bauthor{\bsnm{Dalgaard}, \binits{M.}}:
\batitle{Host responses against the fish parasitizing ciliate ichthyophthirius
  multifiliis}.
\bjtitle{Veterinary Parasitology}
\bvolume{100}(\bissue{1}),
\bfpage{105}--\blpage{116}
(\byear{2001}).
doi:\doiurl{10.1016/S0304-4017(01)00487-3}.
\bcomment{Vaccination and Immunity against Parasites}
\end{barticle}
\endbibitem

%%% 32
\bibitem{Eisen2006}
\begin{barticle}
\bauthor{\bsnm{Eisen}, \binits{J.A.}},
\bauthor{\bsnm{Coyne}, \binits{R.S.}},
\bauthor{\bsnm{Wu}, \binits{M.}},
\bauthor{\bsnm{Wu}, \binits{D.}},
\bauthor{\bsnm{Thiagarajan}, \binits{M.}},
\bauthor{\bsnm{Wortman}, \binits{J.R.}},
\bauthor{\bsnm{Badger}, \binits{J.H.}},
\bauthor{\bsnm{Ren}, \binits{Q.}},
\bauthor{\bsnm{Amedeo}, \binits{P.}},
\bauthor{\bsnm{Jones}, \binits{K.M.}},
\bauthor{\bsnm{Tallon}, \binits{L.J.}},
\bauthor{\bsnm{Delcher}, \binits{A.L.}},
\bauthor{\bsnm{Salzberg}, \binits{S.L.}},
\bauthor{\bsnm{Silva}, \binits{J.C.}},
\bauthor{\bsnm{Haas}, \binits{B.J.}},
\bauthor{\bsnm{Majoros}, \binits{W.H.}},
\bauthor{\bsnm{Farzad}, \binits{M.}},
\bauthor{\bsnm{Carlton}, \binits{J.M.}},
\bauthor{\bsnm{Smith}, \binits{R.K.} \bsuffix{Jr.}},
\bauthor{\bsnm{Garg}, \binits{J.}},
\bauthor{\bsnm{Pearlman}, \binits{R.E.}},
\bauthor{\bsnm{Karrer}, \binits{K.M.}},
\bauthor{\bsnm{Sun}, \binits{L.}},
\bauthor{\bsnm{Manning}, \binits{G.}},
\bauthor{\bsnm{Elde}, \binits{N.C.}},
\bauthor{\bsnm{Turkewitz}, \binits{A.P.}},
\bauthor{\bsnm{Asai}, \binits{D.J.}},
\bauthor{\bsnm{Wilkes}, \binits{D.E.}},
\bauthor{\bsnm{Wang}, \binits{Y.}},
\bauthor{\bsnm{Cai}, \binits{H.}},
\bauthor{\bsnm{Collins}, \binits{K.}},
\bauthor{\bsnm{Stewart}, \binits{B.A.}},
\bauthor{\bsnm{Lee}, \binits{S.R.}},
\bauthor{\bsnm{Wilamowska}, \binits{K.}},
\bauthor{\bsnm{Weinberg}, \binits{Z.}},
\bauthor{\bsnm{Ruzzo}, \binits{W.L.}},
\bauthor{\bsnm{Wloga}, \binits{D.}},
\bauthor{\bsnm{Gaertig}, \binits{J.}},
\bauthor{\bsnm{Frankel}, \binits{J.}},
\bauthor{\bsnm{Tsao}, \binits{C.-C.}},
\bauthor{\bsnm{Gorovsky}, \binits{M.A.}},
\bauthor{\bsnm{Keeling}, \binits{P.J.}},
\bauthor{\bsnm{Waller}, \binits{R.F.}},
\bauthor{\bsnm{Patron}, \binits{N.J.}},
\bauthor{\bsnm{Cherry}, \binits{J.M.}},
\bauthor{\bsnm{Stover}, \binits{N.A.}},
\bauthor{\bsnm{Krieger}, \binits{C.J.}},
\bauthor{\bparticle{del} \bsnm{Toro}, \binits{C.}},
\bauthor{\bsnm{Ryder}, \binits{H.F.}},
\bauthor{\bsnm{Williamson}, \binits{S.C.}},
\bauthor{\bsnm{Barbeau}, \binits{R.A.}},
\bauthor{\bsnm{Hamilton}, \binits{E.P.}},
\bauthor{\bsnm{Orias}, \binits{E.}}:
\batitle{Macronuclear genome sequence of the ciliate tetrahymena thermophila, a
  model eukaryote}.
\bjtitle{PLOS Biology}
\bvolume{4}(\bissue{9}),
\bfpage{1}--\blpage{23}
(\byear{2006}).
doi:\doiurl{10.1371/journal.pbio.0040286}
\end{barticle}
\endbibitem

%%% 33
\bibitem{10.1093/molbev/msz039}
\begin{botherref}
\oauthor{\bsnm{Ashraf}, \binits{K.}},
\oauthor{\bsnm{Nabeel-Shah}, \binits{S.}},
\oauthor{\bsnm{Garg}, \binits{J.}},
\oauthor{\bsnm{Saettone}, \binits{A.}},
\oauthor{\bsnm{Derynck}, \binits{J.}},
\oauthor{\bsnm{Gingras}, \binits{A.-C.}},
\oauthor{\bsnm{Lambert}, \binits{J.-P.}},
\oauthor{\bsnm{Pearlman}, \binits{R.E.}},
\oauthor{\bsnm{Fillingham}, \binits{J.}}:
{Proteomic analysis of histones H2A/H2B and variant Hv1 in Tetrahymena
  thermophila reveals an ancient network of chaperones}.
{Molecular Biology and Evolution}
\textbf{{msz039}}
(2019).
doi:\doiurl{10.1093/molbev/msz039}.
\arxivurl{http://oup.prod.sis.lan/mbe/advance-article-pdf/doi/10.1093/molbev/msz039/27974900/msz039.pdf}
\end{botherref}
\endbibitem

%%% 34
\bibitem{MARTINDALE1982227}
\begin{barticle}
\bauthor{\bsnm{Martindale}, \binits{D.W.}},
\bauthor{\bsnm{Allis}, \binits{C.D.}},
\bauthor{\bsnm{Bruns}, \binits{P.J.}}:
\batitle{Conjugation in tetrahymena thermophila: A temporal analysis of
  cytological stages}.
\bjtitle{Experimental Cell Research}
\bvolume{140}(\bissue{1}),
\bfpage{227}--\blpage{236}
(\byear{1982}).
doi:\doiurl{10.1016/0014-4827(82)90172-0}
\end{barticle}
\endbibitem

%%% 35
\bibitem{Xiong2012}
\begin{barticle}
\bauthor{\bsnm{Xiong}, \binits{J.}},
\bauthor{\bsnm{Lu}, \binits{X.}},
\bauthor{\bsnm{Zhou}, \binits{Z.}},
\bauthor{\bsnm{Chang}, \binits{Y.}},
\bauthor{\bsnm{Yuan}, \binits{D.}},
\bauthor{\bsnm{Tian}, \binits{M.}},
\bauthor{\bsnm{Zhou}, \binits{Z.}},
\bauthor{\bsnm{Wang}, \binits{L.}},
\bauthor{\bsnm{Fu}, \binits{C.}},
\bauthor{\bsnm{Orias}, \binits{E.}},
\bauthor{\bsnm{Miao}, \binits{W.}}:
\batitle{Transcriptome analysis of the model protozoan, tetrahymena
  thermophila, using deep rna sequencing}.
\bjtitle{PLOS ONE}
\bvolume{7}(\bissue{2}),
\bfpage{1}--\blpage{13}
(\byear{2012}).
doi:\doiurl{10.1371/journal.pone.0030630}
\end{barticle}
\endbibitem

%%% 36
\bibitem{10.1371/journal.pcbi.1003326}
\begin{barticle}
\bauthor{\bsnm{Bailey}, \binits{T.}},
\bauthor{\bsnm{Krajewski}, \binits{P.}},
\bauthor{\bsnm{Ladunga}, \binits{I.}},
\bauthor{\bsnm{Lefebvre}, \binits{C.}},
\bauthor{\bsnm{Li}, \binits{Q.}},
\bauthor{\bsnm{Liu}, \binits{T.}},
\bauthor{\bsnm{Madrigal}, \binits{P.}},
\bauthor{\bsnm{Taslim}, \binits{C.}},
\bauthor{\bsnm{Zhang}, \binits{J.}}:
\batitle{{Practical Guidelines for the Comprehensive Analysis of ChIP-seq
  Data}}.
\bjtitle{{PLOS Computational Biology}}
\bvolume{9}(\bissue{11}),
\bfpage{1}--\blpage{8}
(\byear{2013}).
doi:\doiurl{10.1371/journal.pcbi.1003326}
\end{barticle}
\endbibitem

%%% 37
\bibitem{citeulike:11192426}
\begin{barticle}
\bauthor{\bsnm{Landt}, \binits{S.G.}},
\bauthor{\bsnm{Marinov}, \binits{G.K.}},
\bauthor{\bsnm{Kundaje}, \binits{A.}},
\bauthor{\bsnm{Kheradpour}, \binits{P.}},
\bauthor{\bsnm{Pauli}, \binits{F.}},
\bauthor{\bsnm{Batzoglou}, \binits{S.}},
\bauthor{\bsnm{Bernstein}, \binits{B.E.}},
\bauthor{\bsnm{Bickel}, \binits{P.}},
\bauthor{\bsnm{Brown}, \binits{J.B.}},
\bauthor{\bsnm{Cayting}, \binits{P.}},
\bauthor{\bsnm{Chen}, \binits{Y.}},
\bauthor{\bsnm{DeSalvo}, \binits{G.}},
\bauthor{\bsnm{Epstein}, \binits{C.}},
\bauthor{\bsnm{Fisher-Aylor}, \binits{K.I.}},
\bauthor{\bsnm{Euskirchen}, \binits{G.}},
\bauthor{\bsnm{Gerstein}, \binits{M.}},
\bauthor{\bsnm{Gertz}, \binits{J.}},
\bauthor{\bsnm{Hartemink}, \binits{A.J.}},
\bauthor{\bsnm{Hoffman}, \binits{M.M.}},
\bauthor{\bsnm{Iyer}, \binits{V.R.}},
\bauthor{\bsnm{Jung}, \binits{Y.L.}},
\bauthor{\bsnm{Karmakar}, \binits{S.}},
\bauthor{\bsnm{Kellis}, \binits{M.}},
\bauthor{\bsnm{Kharchenko}, \binits{P.V.}},
\bauthor{\bsnm{Li}, \binits{Q.}},
\bauthor{\bsnm{Liu}, \binits{T.}},
\bauthor{\bsnm{Liu}, \binits{X.S.}},
\bauthor{\bsnm{Ma}, \binits{L.}},
\bauthor{\bsnm{Milosavljevic}, \binits{A.}},
\bauthor{\bsnm{Myers}, \binits{R.M.}},
\bauthor{\bsnm{Park}, \binits{P.J.}},
\bauthor{\bsnm{Pazin}, \binits{M.J.}},
\bauthor{\bsnm{Perry}, \binits{M.D.}},
\bauthor{\bsnm{Raha}, \binits{D.}},
\bauthor{\bsnm{Reddy}, \binits{T.E.}},
\bauthor{\bsnm{Rozowsky}, \binits{J.}},
\bauthor{\bsnm{Shoresh}, \binits{N.}},
\bauthor{\bsnm{Sidow}, \binits{A.}},
\bauthor{\bsnm{Slattery}, \binits{M.}},
\bauthor{\bsnm{Stamatoyannopoulos}, \binits{J.A.}},
\bauthor{\bsnm{Tolstorukov}, \binits{M.Y.}},
\bauthor{\bsnm{White}, \binits{K.P.}},
\bauthor{\bsnm{Xi}, \binits{S.}},
\bauthor{\bsnm{Farnham}, \binits{P.J.}},
\bauthor{\bsnm{Lieb}, \binits{J.D.}},
\bauthor{\bsnm{Wold}, \binits{B.J.}},
\bauthor{\bsnm{Snyder}, \binits{M.}}:
\batitle{{ChIP-seq guidelines and practices of the ENCODE and modENCODE
  consortia.}}
\bjtitle{Genome research}
\bvolume{22}(\bissue{9}),
\bfpage{1813}--\blpage{1831}
(\byear{2012}).
doi:\doiurl{10.1101/gr.136184.111}
\end{barticle}
\endbibitem

%%% 38
\bibitem{Cormier2016}
\begin{barticle}
\bauthor{\bsnm{Cormier}, \binits{N.}},
\bauthor{\bsnm{Kolisnik}, \binits{T.}},
\bauthor{\bsnm{Bieda}, \binits{M.}}:
\batitle{{Reusable, extensible, and modifiable R scripts and Kepler workflows
  for comprehensive single set ChIP-seq analysis}}.
\bjtitle{{BMC Bioinformatics}}
\bvolume{17}(\bissue{1}),
\bfpage{270}
(\byear{2016}).
doi:\doiurl{10.1186/s12859-016-1125-3}
\end{barticle}
\endbibitem

%%% 39
\bibitem{Qin2016}
\begin{barticle}
\bauthor{\bsnm{Qin}, \binits{Q.}},
\bauthor{\bsnm{Mei}, \binits{S.}},
\bauthor{\bsnm{Wu}, \binits{Q.}},
\bauthor{\bsnm{Sun}, \binits{H.}},
\bauthor{\bsnm{Li}, \binits{L.}},
\bauthor{\bsnm{Taing}, \binits{L.}},
\bauthor{\bsnm{Chen}, \binits{S.}},
\bauthor{\bsnm{Li}, \binits{F.}},
\bauthor{\bsnm{Liu}, \binits{T.}},
\bauthor{\bsnm{Zang}, \binits{C.}},
\bauthor{\bsnm{Xu}, \binits{H.}},
\bauthor{\bsnm{Chen}, \binits{Y.}},
\bauthor{\bsnm{Meyer}, \binits{C.A.}},
\bauthor{\bsnm{Zhang}, \binits{Y.}},
\bauthor{\bsnm{Brown}, \binits{M.}},
\bauthor{\bsnm{Long}, \binits{H.W.}},
\bauthor{\bsnm{Liu}, \binits{X.S.}}:
\batitle{{ChiLin: a comprehensive ChIP-seq and DNase-seq quality control and
  analysis pipeline}}.
\bjtitle{{BMC Bioinformatics}}
\bvolume{17}(\bissue{1}),
\bfpage{404}
(\byear{2016}).
doi:\doiurl{10.1186/s12859-016-1274-4}
\end{barticle}
\endbibitem

%%% 40
\bibitem{andrews2010fastqc}
\begin{botherref}
\oauthor{\bsnm{Andrews}, \binits{S.}}, et al.:
FastQC: a quality control tool for high throughput sequence data
(2010)
\end{botherref}
\endbibitem

%%% 41
\bibitem{10.1093/bioinformatics/btq033}
\begin{barticle}
\bauthor{\bsnm{Quinlan}, \binits{A.R.}},
\bauthor{\bsnm{Hall}, \binits{I.M.}}:
\batitle{{BEDTools: a flexible suite of utilities for comparing genomic
  features}}.
\bjtitle{Bioinformatics}
\bvolume{26}(\bissue{6}),
\bfpage{841}--\blpage{842}
(\byear{2010}).
doi:\doiurl{10.1093/bioinformatics/btq033}.
\arxivurl{http://oup.prod.sis.lan/bioinformatics/article-pdf/26/6/841/16897802/btq033.pdf}
\end{barticle}
\endbibitem

%%% 42
\bibitem{ashburner2000gene}
\begin{barticle}
\bauthor{\bsnm{Ashburner}, \binits{M.}},
\bauthor{\bsnm{Ball}, \binits{C.A.}},
\bauthor{\bsnm{Blake}, \binits{J.A.}},
\bauthor{\bsnm{Botstein}, \binits{D.}},
\bauthor{\bsnm{Butler}, \binits{H.}},
\bauthor{\bsnm{Cherry}, \binits{J.M.}},
\bauthor{\bsnm{Davis}, \binits{A.P.}},
\bauthor{\bsnm{Dolinski}, \binits{K.}},
\bauthor{\bsnm{Dwight}, \binits{S.S.}},
\bauthor{\bsnm{Eppig}, \binits{J.T.}}, \betal:
\batitle{Gene ontology: tool for the unification of biology}.
\bjtitle{Nature genetics}
\bvolume{25}(\bissue{1}),
\bfpage{25}
(\byear{2000})
\end{barticle}
\endbibitem

%%% 43
\bibitem{wang2017n6}
\begin{barticle}
\bauthor{\bsnm{Wang}, \binits{Y.}},
\bauthor{\bsnm{Chen}, \binits{X.}},
\bauthor{\bsnm{Sheng}, \binits{Y.}},
\bauthor{\bsnm{Liu}, \binits{Y.}},
\bauthor{\bsnm{Gao}, \binits{S.}}:
\batitle{N6-adenine dna methylation is associated with the linker dna of h2a.
  z-containing well-positioned nucleosomes in pol ii-transcribed genes in
  tetrahymena}.
\bjtitle{Nucleic acids research}
\bvolume{45}(\bissue{20}),
\bfpage{11594}--\blpage{11606}
(\byear{2017})
\end{barticle}
\endbibitem

%%% 44
\bibitem{kataoka2015phosphorylation}
\begin{barticle}
\bauthor{\bsnm{Kataoka}, \binits{K.}},
\bauthor{\bsnm{Mochizuki}, \binits{K.}}:
\batitle{Phosphorylation of an hp1-like protein regulates heterochromatin body
  assembly for dna elimination}.
\bjtitle{Developmental cell}
\bvolume{35}(\bissue{6}),
\bfpage{775}--\blpage{788}
(\byear{2015})
\end{barticle}
\endbibitem

\end{thebibliography}

\newcommand{\BMCxmlcomment}[1]{}

\BMCxmlcomment{

<refgrp>

<bibl id="B1">
  <title><p>{SciNet: Lessons Learned from Building a Power-efficient Top-20
  System and Data Centre}</p></title>
  <aug>
    <au><snm>Loken</snm><fnm>C</fnm></au>
    <au><snm>Gruner</snm><fnm>D</fnm></au>
    <au><snm>Groer</snm><fnm>L</fnm></au>
    <au><snm>Peltier</snm><fnm>R</fnm></au>
    <au><snm>Bunn</snm><fnm>N</fnm></au>
    <au><snm>Craig</snm><fnm>M</fnm></au>
    <au><snm>Henriques</snm><fnm>T</fnm></au>
    <au><snm>Dempsey</snm><fnm>J</fnm></au>
    <au><snm>Yu</snm><fnm>CH</fnm></au>
    <au><snm>Chen</snm><fnm>J</fnm></au>
    <au><snm>Dursi</snm><fnm>LJ</fnm></au>
    <au><snm>Chong</snm><fnm>J</fnm></au>
    <au><snm>Northrup</snm><fnm>S</fnm></au>
    <au><snm>Pinto</snm><fnm>J</fnm></au>
    <au><snm>Knecht</snm><fnm>N</fnm></au>
    <au><snm>Zon</snm><fnm>RV</fnm></au>
  </aug>
  <source>Journal of Physics: Conference Series</source>
  <pubdate>2010</pubdate>
  <volume>256</volume>
  <issue>1</issue>
  <fpage>012026</fpage>
  <url>http://stacks.iop.org/1742-6596/256/i=1/a=012026</url>
</bibl>

<bibl id="B2">
  <title><p>Scientific Computing, High-Performance Computing and Data Science
  in Higher Education</p></title>
  <aug>
    <au><snm>Ponce</snm><fnm>M</fnm></au>
    <au><snm>Spence</snm><fnm>E</fnm></au>
    <au><snm>Gruner</snm><fnm>D</fnm></au>
    <au><snm>Zon</snm><fnm>R</fnm></au>
  </aug>
  <source>Journal of Computational Science Education</source>
  <pubdate>2019</pubdate>
</bibl>

<bibl id="B3">
  <title><p>The Sequence of the Human Genome</p></title>
  <aug>
    <au><snm>Venter</snm><fnm>ea</fnm></au>
  </aug>
  <source>Science</source>
  <publisher>American Association for the Advancement of Science</publisher>
  <pubdate>2001</pubdate>
  <volume>291</volume>
  <issue>5507</issue>
  <fpage>1304</fpage>
  <lpage>-1351</lpage>
  <url>http://science.sciencemag.org/content/291/5507/1304</url>
</bibl>

<bibl id="B4">
  <title><p>The NIH Human Microbiome Project</p></title>
  <aug>
    <au><cnm>{The NIH HMP Working Group, Jane Peterson, et al}</cnm></au>
  </aug>
  <source>Genome Research</source>
  <pubdate>2009</pubdate>
  <volume>19</volume>
  <issue>12</issue>
</bibl>

<bibl id="B5">
  <title><p>RNA-Seq: a revolutionary tool for transcriptomics</p></title>
  <aug>
    <au><cnm>{Wang, Zhong; Gerstein, Mark; Snyder, Michael}</cnm></au>
  </aug>
  <source>Genetics</source>
  <pubdate>2009</pubdate>
  <volume>10</volume>
  <issue>1</issue>
</bibl>

<bibl id="B6">
  <title><p>Targeted capture and massively parallel sequencing of 12 human
  exomes</p></title>
  <aug>
    <au><cnm>{Ng, Sarah B; Turner, Emily H; Robertson, Peggy D; Flygare, Steven
  D; Bigham, Abigail W; et al.}</cnm></au>
  </aug>
  <source>Nature</source>
  <pubdate>2009</pubdate>
  <volume>461</volume>
  <issue>7261</issue>
</bibl>

<bibl id="B7">
  <title><p>Next-generation sequencing transforms today's biology</p></title>
  <aug>
    <au><snm>Schuster</snm><fnm>SC</fnm></au>
  </aug>
  <source>Nature Methods</source>
  <pubdate>2008</pubdate>
  <volume>5</volume>
  <url>https://www.nature.com/articles/nmeth1156</url>
</bibl>

<bibl id="B8">
  <title><p>Corrigendum: Performance comparison of benchtop high-throughput
  sequencing platforms</p></title>
  <aug>
    <au><snm>Loman</snm><fnm>NJ</fnm></au>
    <au><snm>Misra</snm><fnm>RV</fnm></au>
    <au><snm>Dallman</snm><fnm>TJ</fnm></au>
    <au><snm>Constantinidou</snm><fnm>C</fnm></au>
    <au><snm>Gharbia</snm><fnm>SE</fnm></au>
    <au><snm>Wain</snm><fnm>J</fnm></au>
    <au><snm>Pallen</snm><fnm>MJ</fnm></au>
  </aug>
  <source>Nature Biotechnology</source>
  <pubdate>2012</pubdate>
  <volume>30</volume>
  <url>http://www.nature.com/articles/nbt0612-562f</url>
</bibl>

<bibl id="B9">
  <title><p>Genome-Wide Mapping of in Vivo Protein-DNA Interactions</p></title>
  <aug>
    <au><snm>Johnson</snm><fnm>DS</fnm></au>
    <au><snm>Mortazavi</snm><fnm>A</fnm></au>
    <au><snm>Myers</snm><fnm>RM</fnm></au>
    <au><snm>Wold</snm><fnm>B</fnm></au>
  </aug>
  <source>Science</source>
  <publisher>American Association for the Advancement of Science</publisher>
  <pubdate>2007</pubdate>
  <volume>316</volume>
  <issue>5830</issue>
  <fpage>1497</fpage>
  <lpage>-1502</lpage>
  <url>http://science.sciencemag.org/content/316/5830/1497</url>
</bibl>

<bibl id="B10">
  <title><p>ChIP–seq: advantages and challenges of a maturing
  technology</p></title>
  <aug>
    <au><snm>Park</snm><fnm>PJ</fnm></au>
  </aug>
  <source>Nature Reviews Genetics</source>
  <pubdate>2009</pubdate>
  <volume>10</volume>
</bibl>

<bibl id="B11">
  <title><p>The impact of next-generation sequencing technology on
  genetics</p></title>
  <aug>
    <au><snm>Mardis</snm><fnm>ER</fnm></au>
  </aug>
  <source>Trends in Genetics</source>
  <pubdate>2008</pubdate>
  <volume>24</volume>
</bibl>

<bibl id="B12">
  <title><p>{Fast and accurate short read alignment with Burrows–Wheeler
  transform}</p></title>
  <aug>
    <au><snm>Li</snm><fnm>H</fnm></au>
    <au><snm>Durbin</snm><fnm>R</fnm></au>
  </aug>
  <source>Bioinformatics</source>
  <pubdate>2009</pubdate>
  <volume>25</volume>
  <issue>14</issue>
  <fpage>1754</fpage>
  <url>http://dx.doi.org/10.1093/bioinformatics/btp324</url>
</bibl>

<bibl id="B13">
  <title><p>{The Sequence Alignment/Map format and SAMtools}</p></title>
  <aug>
    <au><snm>Li</snm><fnm>H</fnm></au>
    <au><snm>Handsaker</snm><fnm>B</fnm></au>
    <au><snm>Wysoker</snm><fnm>A</fnm></au>
    <au><snm>Fennell</snm><fnm>T</fnm></au>
    <au><snm>Ruan</snm><fnm>J</fnm></au>
    <au><snm>Homer</snm><fnm>N</fnm></au>
    <au><snm>Marth</snm><fnm>G</fnm></au>
    <au><snm>Abecasis</snm><fnm>G</fnm></au>
    <au><snm>Durbin</snm><fnm>R</fnm></au>
  </aug>
  <source>Bioinformatics</source>
  <pubdate>2009</pubdate>
  <volume>25</volume>
  <issue>16</issue>
  <fpage>2078</fpage>
  <url>http://dx.doi.org/10.1093/bioinformatics/btp352</url>
</bibl>

<bibl id="B14">
  <title><p>R: A Language and Environment for Statistical Computing</p></title>
  <aug>
    <au><cnm>{R Core Team}</cnm></au>
  </aug>
  <publisher>Vienna, Austria</publisher>
  <pubdate>2016</pubdate>
  <url>https://www.R-project.org/</url>
</bibl>

<bibl id="B15">
  <title><p>CHROMOSOME TERRITORIES, NUCLEAR ARCHITECTURE AND GENE REGULATION IN
  MAMMALIAN CELLS</p></title>
  <aug>
    <au><snm>Cremer</snm><fnm>T.</fnm></au>
    <au><snm>Cremer</snm><fnm>C.</fnm></au>
  </aug>
  <source>Nature Reviews.Genetics</source>
  <pubdate>2001</pubdate>
  <volume>2</volume>
  <issue>4</issue>
  <fpage>292</fpage>
  <lpage>301</lpage>
  <url>http://ezproxy.lib.ryerson.ca/login?url=https://search-proquest-com.ezproxy.lib.ryerson.ca/docview/223749244?accountid=13631</url>
  <note>Copyright - Copyright Nature Publishing Group Apr 2001; Last updated -
  2013-01-27</note>
</bibl>

<bibl id="B16">
  <title><p>Five-Vertebrate ChIP-seq Reveals the Evolutionary Dynamics of
  Transcription Factor Binding</p></title>
  <aug>
    <au><snm>Schmidt</snm><fnm>D</fnm></au>
    <au><snm>Wilson</snm><fnm>MD</fnm></au>
    <au><snm>Ballester</snm><fnm>B</fnm></au>
    <au><snm>Schwalie</snm><fnm>PC</fnm></au>
    <au><snm>Brown</snm><fnm>GD</fnm></au>
    <au><snm>Marshall</snm><fnm>A</fnm></au>
    <au><snm>Kutter</snm><fnm>C</fnm></au>
    <au><snm>Watt</snm><fnm>S</fnm></au>
    <au><snm>Martinez Jimenez</snm><fnm>CP</fnm></au>
    <au><snm>Mackay</snm><fnm>S</fnm></au>
    <au><snm>Talianidis</snm><fnm>I</fnm></au>
    <au><snm>Flicek</snm><fnm>P</fnm></au>
    <au><snm>Odom</snm><fnm>DT</fnm></au>
  </aug>
  <source>Science</source>
  <pubdate>2010</pubdate>
  <volume>328</volume>
  <issue>5981</issue>
  <fpage>1036</fpage>
  <lpage>-1040</lpage>
  <url>http://science.sciencemag.org/content/328/5981/1036</url>
</bibl>

<bibl id="B17">
  <title><p>Model-based Analysis of ChIP-Seq (MACS)</p></title>
  <aug>
    <au><snm>Zhang</snm><fnm>Y</fnm></au>
    <au><snm>Liu</snm><fnm>T</fnm></au>
    <au><snm>Meyer</snm><fnm>CA</fnm></au>
    <au><snm>Eeckhoute</snm><fnm>J</fnm></au>
    <au><snm>Johnson</snm><fnm>DS</fnm></au>
    <au><snm>Bernstein</snm><fnm>BE</fnm></au>
    <au><snm>Nusbaum</snm><fnm>C</fnm></au>
    <au><snm>Myers</snm><fnm>RM</fnm></au>
    <au><snm>Brown</snm><fnm>M</fnm></au>
    <au><snm>Li</snm><fnm>W</fnm></au>
    <au><snm>Liu</snm><fnm>XS</fnm></au>
  </aug>
  <source>Genome Biology</source>
  <pubdate>2008</pubdate>
  <volume>9</volume>
  <issue>9</issue>
  <fpage>R137</fpage>
  <url>https://doi.org/10.1186/gb-2008-9-9-r137</url>
</bibl>

<bibl id="B18">
  <title><p>Tetrahymena Genome Database (TGD): a new genomic resource for
  Tetrahymena thermophila research</p></title>
  <aug>
    <au><snm>Stover</snm><fnm>NA</fnm></au>
    <au><snm>Krieger</snm><fnm>CJ</fnm></au>
    <au><snm>Binkley</snm><fnm>G</fnm></au>
    <au><snm>Dong</snm><fnm>Q</fnm></au>
    <au><snm>Fisk</snm><fnm>DG</fnm></au>
    <au><snm>Nash</snm><fnm>R</fnm></au>
    <au><snm>Sethuraman</snm><fnm>A</fnm></au>
    <au><snm>Weng</snm><fnm>S</fnm></au>
    <au><snm>Cherry</snm><fnm>JM</fnm></au>
  </aug>
  <source>Nucleic Acids Research</source>
  <pubdate>2006</pubdate>
  <volume>34</volume>
  <issue>suppl\_1</issue>
  <fpage>D500</fpage>
  <lpage>D503</lpage>
  <url>http://dx.doi.org/10.1093/nar/gkj054</url>
</bibl>

<bibl id="B19">
  <title><p>PBS: A Unified Priority-based Scheduler</p></title>
  <aug>
    <au><snm>Feng</snm><fnm>H</fnm></au>
    <au><snm>Misra</snm><fnm>V</fnm></au>
    <au><snm>Rubenstein</snm><fnm>D</fnm></au>
  </aug>
  <source>SIGMETRICS Perform. Eval. Rev.</source>
  <publisher>New York, NY, USA: ACM</publisher>
  <pubdate>2007</pubdate>
  <volume>35</volume>
  <issue>1</issue>
  <fpage>203</fpage>
  <lpage>-214</lpage>
  <url>http://doi.acm.org/10.1145/1269899.1254906</url>
</bibl>

<bibl id="B20">
  <title><p>A Priority-Based Budget Scheduler with Conservative Dataflow
  Model</p></title>
  <aug>
    <au><snm>Steine</snm><fnm>M</fnm></au>
    <au><snm>Bekooij</snm><fnm>M</fnm></au>
    <au><snm>Wiggers</snm><fnm>M</fnm></au>
  </aug>
  <source>Proceedings of the 2009 12th Euromicro Conference on Digital System
  Design, Architectures, Methods and Tools</source>
  <publisher>Washington, DC, USA: IEEE Computer Society</publisher>
  <series><title><p>DSD '09</p></title></series>
  <pubdate>2009</pubdate>
  <fpage>37</fpage>
  <lpage>-44</lpage>
  <url>https://doi.org/10.1109/DSD.2009.148</url>
</bibl>

<bibl id="B21">
  <title><p>Best Practices for the Deployment and Management of Production HPC
  Clusters</p></title>
  <aug>
    <au><snm>McLay</snm><fnm>R</fnm></au>
    <au><snm>Schulz</snm><fnm>KW</fnm></au>
    <au><snm>Barth</snm><fnm>WL</fnm></au>
    <au><snm>Minyard</snm><fnm>T</fnm></au>
  </aug>
  <source>State of the Practice Reports</source>
  <publisher>New York, NY, USA: ACM</publisher>
  <series><title><p>SC '11</p></title></series>
  <pubdate>2011</pubdate>
  <fpage>9:1</fpage>
  <lpage>-9:11</lpage>
  <url>http://doi.acm.org/10.1145/2063348.2063360</url>
</bibl>

<bibl id="B22">
  <title><p>SLURM: Simple Linux Utility for Resource Management</p></title>
  <aug>
    <au><snm>Yoo</snm><fnm>AB</fnm></au>
    <au><snm>Jette</snm><fnm>MA</fnm></au>
    <au><snm>Grondona</snm><fnm>M</fnm></au>
  </aug>
  <source>Job Scheduling Strategies for Parallel Processing</source>
  <publisher>Berlin, Heidelberg: Springer Berlin Heidelberg</publisher>
  <editor>Feitelson, Dror and Rudolph, Larry and Schwiegelshohn, Uwe</editor>
  <pubdate>2003</pubdate>
  <fpage>44</fpage>
  <lpage>-60</lpage>
</bibl>

<bibl id="B23">
  <title><p>The bromodomain-containing protein Ibd1 links multiple
  chromatin-related protein complexes to highly expressed genes in Tetrahymena
  thermophila</p></title>
  <aug>
    <au><snm>Saettone</snm><fnm>A</fnm></au>
    <au><snm>Garg</snm><fnm>J</fnm></au>
    <au><snm>Lambert</snm><fnm>JP</fnm></au>
    <au><snm>Nabeel Shah</snm><fnm>S</fnm></au>
    <au><snm>Ponce</snm><fnm>M</fnm></au>
    <au><snm>Burtch</snm><fnm>A</fnm></au>
    <au><snm>Thuppu Mudalige</snm><fnm>C</fnm></au>
    <au><snm>Gingras</snm><fnm>AC</fnm></au>
    <au><snm>Pearlman</snm><fnm>RE</fnm></au>
    <au><snm>Fillingham</snm><fnm>J</fnm></au>
  </aug>
  <source>Epigenetics {\&} Chromatin</source>
  <pubdate>2018</pubdate>
  <volume>11</volume>
  <issue>1</issue>
  <fpage>10</fpage>
  <url>https://doi.org/10.1186/s13072-018-0180-6</url>
</bibl>

<bibl id="B24">
  <title><p>{Integrative Genomics Viewer}</p></title>
  <aug>
    <au><snm>Robinson</snm><fnm>JT</fnm></au>
    <au><snm>Thorvaldsd{\'o}ttir</snm><fnm>H</fnm></au>
    <au><snm>Winckler</snm><fnm>W</fnm></au>
    <au><snm>Guttman</snm><fnm>M</fnm></au>
    <au><snm>Lander</snm><fnm>ES</fnm></au>
    <au><snm>Getz</snm><fnm>G</fnm></au>
    <au><snm>Mesirov</snm><fnm>JP</fnm></au>
  </aug>
  <source>{Nature Biotechnology}</source>
  <publisher>Nature Research</publisher>
  <pubdate>2011</pubdate>
  <volume>29</volume>
  <issue>1</issue>
  <fpage>24</fpage>
  <lpage>-26</lpage>
</bibl>

<bibl id="B25">
  <title><p>{Deploying a Top-100 Supercomputer for Large Parallel Workloads:
  the Niagara Supercomputer}</p></title>
  <aug>
    <au><cnm>{Ponce et al.}</cnm></au>
  </aug>
  <source>{in prep.}</source>
  <pubdate>2019</pubdate>
</bibl>

<bibl id="B26">
  <title><p>Sambamba: fast processing of NGS alignment formats</p></title>
  <aug>
    <au><snm>Tarasov</snm><fnm>A</fnm></au>
    <au><snm>Vilella</snm><fnm>AJ</fnm></au>
    <au><snm>Cuppen</snm><fnm>E</fnm></au>
    <au><snm>Nijman</snm><fnm>IJ</fnm></au>
    <au><snm>Prins</snm><fnm>P</fnm></au>
  </aug>
  <source>Bioinformatics</source>
  <pubdate>2015</pubdate>
  <volume>31</volume>
  <issue>12</issue>
  <fpage>2032</fpage>
  <lpage>2034</lpage>
  <url>http://dx.doi.org/10.1093/bioinformatics/btv098</url>
</bibl>

<bibl id="B27">
  <title><p>The circumsporozoite protein of Plasmodium falciparum is expressed
  and localized to the cell surface in the free-living ciliate Tetrahymena
  thermophila</p></title>
  <aug>
    <au><snm>Peterson</snm><fnm>DS</fnm></au>
    <au><snm>Gao</snm><fnm>Y</fnm></au>
    <au><snm>Asokan</snm><fnm>K</fnm></au>
    <au><snm>Gaertig</snm><fnm>J</fnm></au>
  </aug>
  <source>Molecular and Biochemical Parasitology</source>
  <pubdate>2002</pubdate>
  <volume>122</volume>
  <issue>2</issue>
  <fpage>119</fpage>
  <lpage>126</lpage>
  <url>http://www.sciencedirect.com/science/article/pii/S0166685102000798</url>
</bibl>

<bibl id="B28">
  <title><p>Guanylyl cyclases with the topology of mammalian adenylyl cyclases
  and an N-terminal P-type ATPase-like domain in Paramecium, Tetrahymena and
  Plasmodium</p></title>
  <aug>
    <au><snm>Linder</snm><fnm>JU</fnm></au>
    <au><snm>Engel</snm><fnm>P</fnm></au>
    <au><snm>Reimer</snm><fnm>A</fnm></au>
    <au><snm>Kr{\"u}ger</snm><fnm>T</fnm></au>
    <au><snm>Plattner</snm><fnm>H</fnm></au>
    <au><snm>Schultz</snm><fnm>A</fnm></au>
    <au><snm>Schultz</snm><fnm>JE</fnm></au>
  </aug>
  <source>The EMBO Journal</source>
  <publisher>EMBO Press</publisher>
  <pubdate>1999</pubdate>
  <volume>18</volume>
  <issue>15</issue>
  <fpage>4222</fpage>
  <lpage>-4232</lpage>
  <url>http://emboj.embopress.org/content/18/15/4222</url>
</bibl>

<bibl id="B29">
  <title><p>cGMP signalling: different ways to create a pathway</p></title>
  <aug>
    <au><snm>Roelofs</snm><fnm>J</fnm></au>
    <au><snm>Smith</snm><fnm>JL</fnm></au>
    <au><snm>Haastert</snm><fnm>PJV</fnm></au>
  </aug>
  <source>Trends in Genetics</source>
  <pubdate>2003</pubdate>
  <volume>19</volume>
  <issue>3</issue>
  <fpage>132</fpage>
  <lpage>134</lpage>
  <url>http://www.sciencedirect.com/science/article/pii/S0168952502000446</url>
</bibl>

<bibl id="B30">
  <title><p>Morphological Variation and Phylogenetic Analysis of the
  Dinoflagellate Gymnodinium aureolum from a Tributary of Chesapeake
  Bay</p></title>
  <aug>
    <au><snm>Tang</snm><fnm>YZ</fnm></au>
    <au><snm>Egerton</snm><fnm>TA</fnm></au>
    <au><snm>Kong</snm><fnm>L</fnm></au>
    <au><snm>Marshall</snm><fnm>HG</fnm></au>
  </aug>
  <source>Journal of Eukaryotic Microbiology</source>
  <publisher>Blackwell Publishing Inc</publisher>
  <pubdate>2008</pubdate>
  <volume>55</volume>
  <issue>2</issue>
  <fpage>91</fpage>
  <lpage>-99</lpage>
  <url>http://dx.doi.org/10.1111/j.1550-7408.2008.00305.x</url>
</bibl>

<bibl id="B31">
  <title><p>Host responses against the fish parasitizing ciliate
  Ichthyophthirius multifiliis</p></title>
  <aug>
    <au><snm>Buchmann</snm><fnm>K</fnm></au>
    <au><snm>Sigh</snm><fnm>J</fnm></au>
    <au><snm>Nielsen</snm><fnm>C.V</fnm></au>
    <au><snm>Dalgaard</snm><fnm>M</fnm></au>
  </aug>
  <source>Veterinary Parasitology</source>
  <pubdate>2001</pubdate>
  <volume>100</volume>
  <issue>1</issue>
  <fpage>105</fpage>
  <lpage>116</lpage>
  <url>http://www.sciencedirect.com/science/article/pii/S0304401701004873</url>
  <note>Vaccination and Immunity against Parasites</note>
</bibl>

<bibl id="B32">
  <title><p>Macronuclear Genome Sequence of the Ciliate Tetrahymena
  thermophila, a Model Eukaryote</p></title>
  <aug>
    <au><snm>Eisen</snm><fnm>JA</fnm></au>
    <au><snm>Coyne</snm><fnm>RS</fnm></au>
    <au><snm>Wu</snm><fnm>M</fnm></au>
    <au><snm>Wu</snm><fnm>D</fnm></au>
    <au><snm>Thiagarajan</snm><fnm>M</fnm></au>
    <au><snm>Wortman</snm><fnm>JR</fnm></au>
    <au><snm>Badger</snm><fnm>JH</fnm></au>
    <au><snm>Ren</snm><fnm>Q</fnm></au>
    <au><snm>Amedeo</snm><fnm>P</fnm></au>
    <au><snm>Jones</snm><fnm>KM</fnm></au>
    <au><snm>Tallon</snm><fnm>LJ</fnm></au>
    <au><snm>Delcher</snm><fnm>AL</fnm></au>
    <au><snm>Salzberg</snm><fnm>SL</fnm></au>
    <au><snm>Silva</snm><fnm>JC</fnm></au>
    <au><snm>Haas</snm><fnm>BJ</fnm></au>
    <au><snm>Majoros</snm><fnm>WH</fnm></au>
    <au><snm>Farzad</snm><fnm>M</fnm></au>
    <au><snm>Carlton</snm><fnm>JM</fnm></au>
    <au><snm>Smith</snm><fnm>RK</fnm></au>
    <au><snm>Garg</snm><fnm>J</fnm></au>
    <au><snm>Pearlman</snm><fnm>RE</fnm></au>
    <au><snm>Karrer</snm><fnm>KM</fnm></au>
    <au><snm>Sun</snm><fnm>L</fnm></au>
    <au><snm>Manning</snm><fnm>G</fnm></au>
    <au><snm>Elde</snm><fnm>NC</fnm></au>
    <au><snm>Turkewitz</snm><fnm>AP</fnm></au>
    <au><snm>Asai</snm><fnm>DJ</fnm></au>
    <au><snm>Wilkes</snm><fnm>DE</fnm></au>
    <au><snm>Wang</snm><fnm>Y</fnm></au>
    <au><snm>Cai</snm><fnm>H</fnm></au>
    <au><snm>Collins</snm><fnm>K</fnm></au>
    <au><snm>Stewart</snm><fnm>BA</fnm></au>
    <au><snm>Lee</snm><fnm>SR</fnm></au>
    <au><snm>Wilamowska</snm><fnm>K</fnm></au>
    <au><snm>Weinberg</snm><fnm>Z</fnm></au>
    <au><snm>Ruzzo</snm><fnm>WL</fnm></au>
    <au><snm>Wloga</snm><fnm>D</fnm></au>
    <au><snm>Gaertig</snm><fnm>J</fnm></au>
    <au><snm>Frankel</snm><fnm>J</fnm></au>
    <au><snm>Tsao</snm><fnm>CC</fnm></au>
    <au><snm>Gorovsky</snm><fnm>MA</fnm></au>
    <au><snm>Keeling</snm><fnm>PJ</fnm></au>
    <au><snm>Waller</snm><fnm>RF</fnm></au>
    <au><snm>Patron</snm><fnm>NJ</fnm></au>
    <au><snm>Cherry</snm><fnm>JM</fnm></au>
    <au><snm>Stover</snm><fnm>NA</fnm></au>
    <au><snm>Krieger</snm><fnm>CJ</fnm></au>
    <au><snm>Toro</snm><fnm>C</fnm></au>
    <au><snm>Ryder</snm><fnm>HF</fnm></au>
    <au><snm>Williamson</snm><fnm>SC</fnm></au>
    <au><snm>Barbeau</snm><fnm>RA</fnm></au>
    <au><snm>Hamilton</snm><fnm>EP</fnm></au>
    <au><snm>Orias</snm><fnm>E</fnm></au>
  </aug>
  <source>PLOS Biology</source>
  <publisher>Public Library of Science</publisher>
  <pubdate>2006</pubdate>
  <volume>4</volume>
  <issue>9</issue>
  <fpage>1</fpage>
  <lpage>23</lpage>
  <url>https://doi.org/10.1371/journal.pbio.0040286</url>
</bibl>

<bibl id="B33">
  <title><p>{Proteomic analysis of histones H2A/H2B and variant Hv1 in
  Tetrahymena thermophila reveals an ancient network of chaperones}</p></title>
  <aug>
    <au><snm>Ashraf</snm><fnm>K</fnm></au>
    <au><snm>Nabeel Shah</snm><fnm>S</fnm></au>
    <au><snm>Garg</snm><fnm>J</fnm></au>
    <au><snm>Saettone</snm><fnm>A</fnm></au>
    <au><snm>Derynck</snm><fnm>J</fnm></au>
    <au><snm>Gingras</snm><fnm>AC</fnm></au>
    <au><snm>Lambert</snm><fnm>JP</fnm></au>
    <au><snm>Pearlman</snm><fnm>RE</fnm></au>
    <au><snm>Fillingham</snm><fnm>J</fnm></au>
  </aug>
  <source>{Molecular Biology and Evolution}</source>
  <pubdate>2019</pubdate>
  <volume>{msz039}</volume>
  <url>https://dx.doi.org/10.1093/molbev/msz039</url>
</bibl>

<bibl id="B34">
  <title><p>Conjugation in Tetrahymena thermophila: A temporal analysis of
  cytological stages</p></title>
  <aug>
    <au><snm>Martindale</snm><fnm>DW</fnm></au>
    <au><snm>Allis</snm><fnm>C</fnm></au>
    <au><snm>Bruns</snm><fnm>PJ</fnm></au>
  </aug>
  <source>Experimental Cell Research</source>
  <pubdate>1982</pubdate>
  <volume>140</volume>
  <issue>1</issue>
  <fpage>227</fpage>
  <lpage>236</lpage>
  <url>http://www.sciencedirect.com/science/article/pii/0014482782901720</url>
</bibl>

<bibl id="B35">
  <title><p>Transcriptome Analysis of the Model Protozoan, Tetrahymena
  thermophila, Using Deep RNA Sequencing</p></title>
  <aug>
    <au><snm>Xiong</snm><fnm>J</fnm></au>
    <au><snm>Lu</snm><fnm>X</fnm></au>
    <au><snm>Zhou</snm><fnm>Z</fnm></au>
    <au><snm>Chang</snm><fnm>Y</fnm></au>
    <au><snm>Yuan</snm><fnm>D</fnm></au>
    <au><snm>Tian</snm><fnm>M</fnm></au>
    <au><snm>Zhou</snm><fnm>Z</fnm></au>
    <au><snm>Wang</snm><fnm>L</fnm></au>
    <au><snm>Fu</snm><fnm>C</fnm></au>
    <au><snm>Orias</snm><fnm>E</fnm></au>
    <au><snm>Miao</snm><fnm>W</fnm></au>
  </aug>
  <source>PLOS ONE</source>
  <publisher>Public Library of Science</publisher>
  <pubdate>2012</pubdate>
  <volume>7</volume>
  <issue>2</issue>
  <fpage>1</fpage>
  <lpage>13</lpage>
  <url>https://doi.org/10.1371/journal.pone.0030630</url>
</bibl>

<bibl id="B36">
  <title><p>{Practical Guidelines for the Comprehensive Analysis of ChIP-seq
  Data}</p></title>
  <aug>
    <au><snm>Bailey</snm><fnm>T</fnm></au>
    <au><snm>Krajewski</snm><fnm>P</fnm></au>
    <au><snm>Ladunga</snm><fnm>I</fnm></au>
    <au><snm>Lefebvre</snm><fnm>C</fnm></au>
    <au><snm>Li</snm><fnm>Q</fnm></au>
    <au><snm>Liu</snm><fnm>T</fnm></au>
    <au><snm>Madrigal</snm><fnm>P</fnm></au>
    <au><snm>Taslim</snm><fnm>C</fnm></au>
    <au><snm>Zhang</snm><fnm>J</fnm></au>
  </aug>
  <source>{PLOS Computational Biology}</source>
  <publisher>Public Library of Science</publisher>
  <pubdate>2013</pubdate>
  <volume>9</volume>
  <issue>11</issue>
  <fpage>1</fpage>
  <lpage>8</lpage>
  <url>https://doi.org/10.1371/journal.pcbi.1003326</url>
</bibl>

<bibl id="B37">
  <title><p>{ChIP-seq guidelines and practices of the ENCODE and modENCODE
  consortia.}</p></title>
  <aug>
    <au><snm>Landt</snm><fnm>SG</fnm></au>
    <au><snm>Marinov</snm><fnm>GK</fnm></au>
    <au><snm>Kundaje</snm><fnm>A</fnm></au>
    <au><snm>Kheradpour</snm><fnm>P</fnm></au>
    <au><snm>Pauli</snm><fnm>F</fnm></au>
    <au><snm>Batzoglou</snm><fnm>S</fnm></au>
    <au><snm>Bernstein</snm><fnm>BE</fnm></au>
    <au><snm>Bickel</snm><fnm>P</fnm></au>
    <au><snm>Brown</snm><fnm>JB</fnm></au>
    <au><snm>Cayting</snm><fnm>P</fnm></au>
    <au><snm>Chen</snm><fnm>Y</fnm></au>
    <au><snm>DeSalvo</snm><fnm>G</fnm></au>
    <au><snm>Epstein</snm><fnm>C</fnm></au>
    <au><snm>Fisher Aylor</snm><fnm>KI</fnm></au>
    <au><snm>Euskirchen</snm><fnm>G</fnm></au>
    <au><snm>Gerstein</snm><fnm>M</fnm></au>
    <au><snm>Gertz</snm><fnm>J</fnm></au>
    <au><snm>Hartemink</snm><fnm>AJ</fnm></au>
    <au><snm>Hoffman</snm><fnm>MM</fnm></au>
    <au><snm>Iyer</snm><fnm>VR</fnm></au>
    <au><snm>Jung</snm><fnm>YL</fnm></au>
    <au><snm>Karmakar</snm><fnm>S</fnm></au>
    <au><snm>Kellis</snm><fnm>M</fnm></au>
    <au><snm>Kharchenko</snm><fnm>PV</fnm></au>
    <au><snm>Li</snm><fnm>Q</fnm></au>
    <au><snm>Liu</snm><fnm>T</fnm></au>
    <au><snm>Liu</snm><fnm>XS</fnm></au>
    <au><snm>Ma</snm><fnm>L</fnm></au>
    <au><snm>Milosavljevic</snm><fnm>A</fnm></au>
    <au><snm>Myers</snm><fnm>RM</fnm></au>
    <au><snm>Park</snm><fnm>PJ</fnm></au>
    <au><snm>Pazin</snm><fnm>MJ</fnm></au>
    <au><snm>Perry</snm><fnm>MD</fnm></au>
    <au><snm>Raha</snm><fnm>D</fnm></au>
    <au><snm>Reddy</snm><fnm>TE</fnm></au>
    <au><snm>Rozowsky</snm><fnm>J</fnm></au>
    <au><snm>Shoresh</snm><fnm>N</fnm></au>
    <au><snm>Sidow</snm><fnm>A</fnm></au>
    <au><snm>Slattery</snm><fnm>M</fnm></au>
    <au><snm>Stamatoyannopoulos</snm><fnm>JA</fnm></au>
    <au><snm>Tolstorukov</snm><fnm>MY</fnm></au>
    <au><snm>White</snm><fnm>KP</fnm></au>
    <au><snm>Xi</snm><fnm>S</fnm></au>
    <au><snm>Farnham</snm><fnm>PJ</fnm></au>
    <au><snm>Lieb</snm><fnm>JD</fnm></au>
    <au><snm>Wold</snm><fnm>BJ</fnm></au>
    <au><snm>Snyder</snm><fnm>M</fnm></au>
  </aug>
  <source>Genome research</source>
  <publisher>Cold Spring Harbor Laboratory Press</publisher>
  <pubdate>2012</pubdate>
  <volume>22</volume>
  <issue>9</issue>
  <fpage>1813</fpage>
  <lpage>-1831</lpage>
  <url>http://dx.doi.org/10.1101/gr.136184.111</url>
</bibl>

<bibl id="B38">
  <title><p>{Reusable, extensible, and modifiable R scripts and Kepler
  workflows for comprehensive single set ChIP-seq analysis}</p></title>
  <aug>
    <au><snm>Cormier</snm><fnm>N</fnm></au>
    <au><snm>Kolisnik</snm><fnm>T</fnm></au>
    <au><snm>Bieda</snm><fnm>M</fnm></au>
  </aug>
  <source>{BMC Bioinformatics}</source>
  <pubdate>2016</pubdate>
  <volume>17</volume>
  <issue>1</issue>
  <fpage>270</fpage>
  <url>http://dx.doi.org/10.1186/s12859-016-1125-3</url>
</bibl>

<bibl id="B39">
  <title><p>{ChiLin: a comprehensive ChIP-seq and DNase-seq quality control and
  analysis pipeline}</p></title>
  <aug>
    <au><snm>Qin</snm><fnm>Q</fnm></au>
    <au><snm>Mei</snm><fnm>S</fnm></au>
    <au><snm>Wu</snm><fnm>Q</fnm></au>
    <au><snm>Sun</snm><fnm>H</fnm></au>
    <au><snm>Li</snm><fnm>L</fnm></au>
    <au><snm>Taing</snm><fnm>L</fnm></au>
    <au><snm>Chen</snm><fnm>S</fnm></au>
    <au><snm>Li</snm><fnm>F</fnm></au>
    <au><snm>Liu</snm><fnm>T</fnm></au>
    <au><snm>Zang</snm><fnm>C</fnm></au>
    <au><snm>Xu</snm><fnm>H</fnm></au>
    <au><snm>Chen</snm><fnm>Y</fnm></au>
    <au><snm>Meyer</snm><fnm>CA</fnm></au>
    <au><snm>Zhang</snm><fnm>Y</fnm></au>
    <au><snm>Brown</snm><fnm>M</fnm></au>
    <au><snm>Long</snm><fnm>HW</fnm></au>
    <au><snm>Liu</snm><fnm>XS</fnm></au>
  </aug>
  <source>{BMC Bioinformatics}</source>
  <pubdate>2016</pubdate>
  <volume>17</volume>
  <issue>1</issue>
  <fpage>404</fpage>
  <url>http://dx.doi.org/10.1186/s12859-016-1274-4</url>
</bibl>

<bibl id="B40">
  <title><p>FastQC: a quality control tool for high throughput sequence
  data</p></title>
  <aug>
    <au><snm>Andrews</snm><fnm>S</fnm></au>
    <au><cnm>others</cnm></au>
  </aug>
  <pubdate>2010</pubdate>
</bibl>

<bibl id="B41">
  <title><p>{BEDTools: a flexible suite of utilities for comparing genomic
  features}</p></title>
  <aug>
    <au><snm>Quinlan</snm><fnm>AR</fnm></au>
    <au><snm>Hall</snm><fnm>IM</fnm></au>
  </aug>
  <source>Bioinformatics</source>
  <pubdate>2010</pubdate>
  <volume>26</volume>
  <issue>6</issue>
  <fpage>841</fpage>
  <lpage>842</lpage>
  <url>https://dx.doi.org/10.1093/bioinformatics/btq033</url>
</bibl>

<bibl id="B42">
  <title><p>Gene Ontology: tool for the unification of biology</p></title>
  <aug>
    <au><snm>Ashburner</snm><fnm>M</fnm></au>
    <au><snm>Ball</snm><fnm>CA</fnm></au>
    <au><snm>Blake</snm><fnm>JA</fnm></au>
    <au><snm>Botstein</snm><fnm>D</fnm></au>
    <au><snm>Butler</snm><fnm>H</fnm></au>
    <au><snm>Cherry</snm><fnm>JM</fnm></au>
    <au><snm>Davis</snm><fnm>AP</fnm></au>
    <au><snm>Dolinski</snm><fnm>K</fnm></au>
    <au><snm>Dwight</snm><fnm>SS</fnm></au>
    <au><snm>Eppig</snm><fnm>JT</fnm></au>
    <au><cnm>others</cnm></au>
  </aug>
  <source>Nature genetics</source>
  <publisher>Nature Publishing Group</publisher>
  <pubdate>2000</pubdate>
  <volume>25</volume>
  <issue>1</issue>
  <fpage>25</fpage>
</bibl>

<bibl id="B43">
  <title><p>N6-adenine DNA methylation is associated with the linker DNA of
  H2A. Z-containing well-positioned nucleosomes in Pol II-transcribed genes in
  Tetrahymena</p></title>
  <aug>
    <au><snm>Wang</snm><fnm>Y</fnm></au>
    <au><snm>Chen</snm><fnm>X</fnm></au>
    <au><snm>Sheng</snm><fnm>Y</fnm></au>
    <au><snm>Liu</snm><fnm>Y</fnm></au>
    <au><snm>Gao</snm><fnm>S</fnm></au>
  </aug>
  <source>Nucleic acids research</source>
  <publisher>Oxford University Press</publisher>
  <pubdate>2017</pubdate>
  <volume>45</volume>
  <issue>20</issue>
  <fpage>11594</fpage>
  <lpage>-11606</lpage>
</bibl>

<bibl id="B44">
  <title><p>Phosphorylation of an HP1-like protein regulates heterochromatin
  body assembly for DNA elimination</p></title>
  <aug>
    <au><snm>Kataoka</snm><fnm>K</fnm></au>
    <au><snm>Mochizuki</snm><fnm>K</fnm></au>
  </aug>
  <source>Developmental cell</source>
  <publisher>Elsevier</publisher>
  <pubdate>2015</pubdate>
  <volume>35</volume>
  <issue>6</issue>
  <fpage>775</fpage>
  <lpage>-788</lpage>
</bibl>

</refgrp>
} % end of \BMCxmlcomment

\clearpage
\section{Figures}

  \begin{figure}[h!]
	\includegraphics[width=\columnwidth]{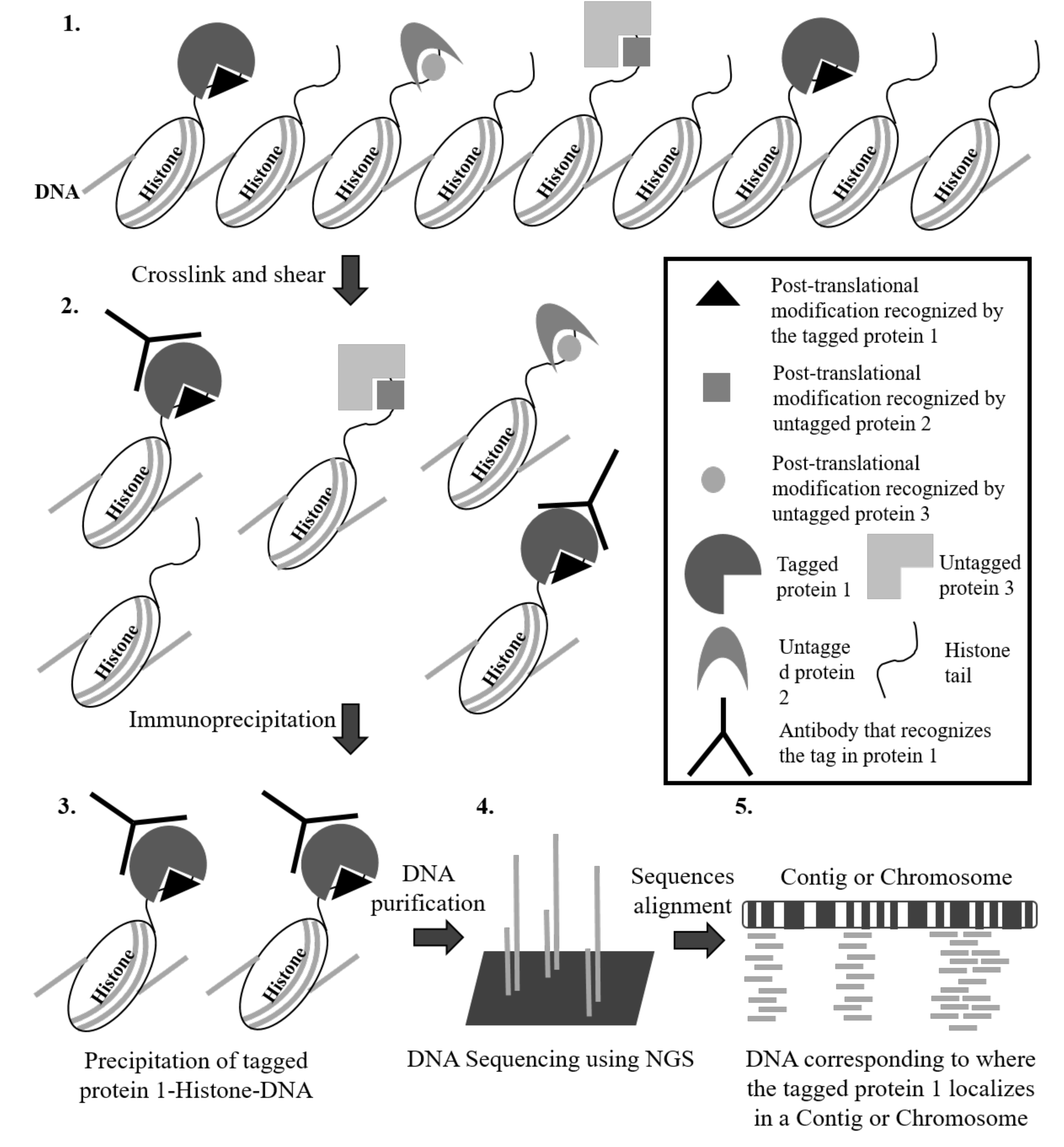}
  \caption{Diagram summarizing the \textit{ChIP-Seq} technique used to prepare
                the samples and generate the data from the ``wet-lab'':
		1) Native state of chromatin.
		2) Specific antibodies recognize the tagged proteins.
                3) Isolation of tagged protein plus its interacting chromatin.
                4) After DNA purification and library preparation NGS is performed.
		5) The output data from NGS is aligned to \textit{Tetrahymena thermophila}'s genome assembly.}
  \label{fig:ChIP}
      \end{figure}

\begin{figure}[h!]
	\includegraphics[width=\columnwidth]{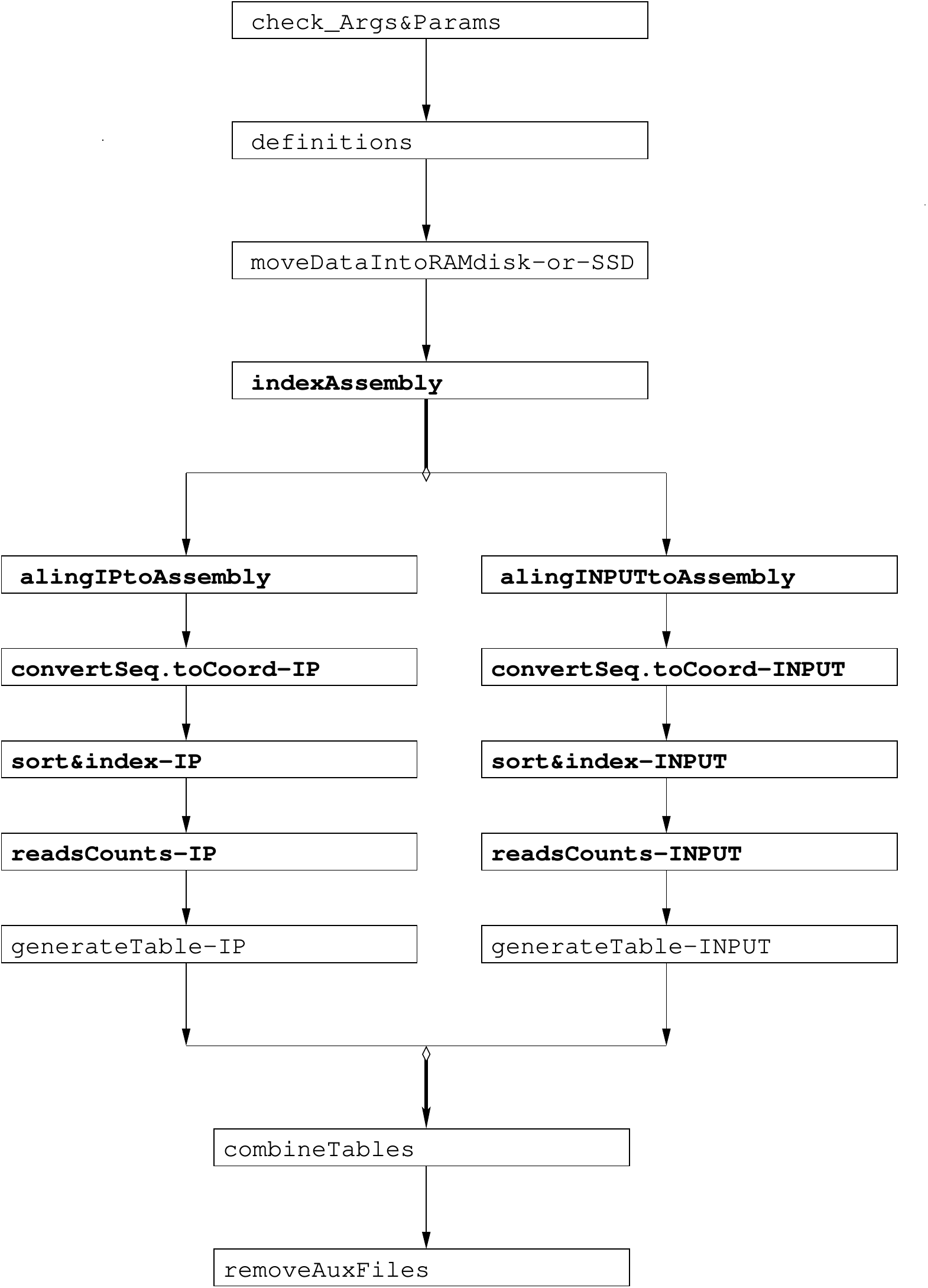}	
        \caption{Schematic diagram of the tasks implemented 
        for the RACS \textit{core pipeline}. Included are details of the processing 
        stages in relation to the scaffold based genome and the breakdown of 
        each these steps. 
        Bold names, indicate bioinformatic specific modules
        while normal fonts represent generic ones.
        The bifurcation represents tasks that can be executed in parallel, 
        as there is no data dependency among them.}
        \label{fig:Parallel}
\end{figure}
    \begin{figure}[h!]
	\includegraphics[width=\columnwidth]{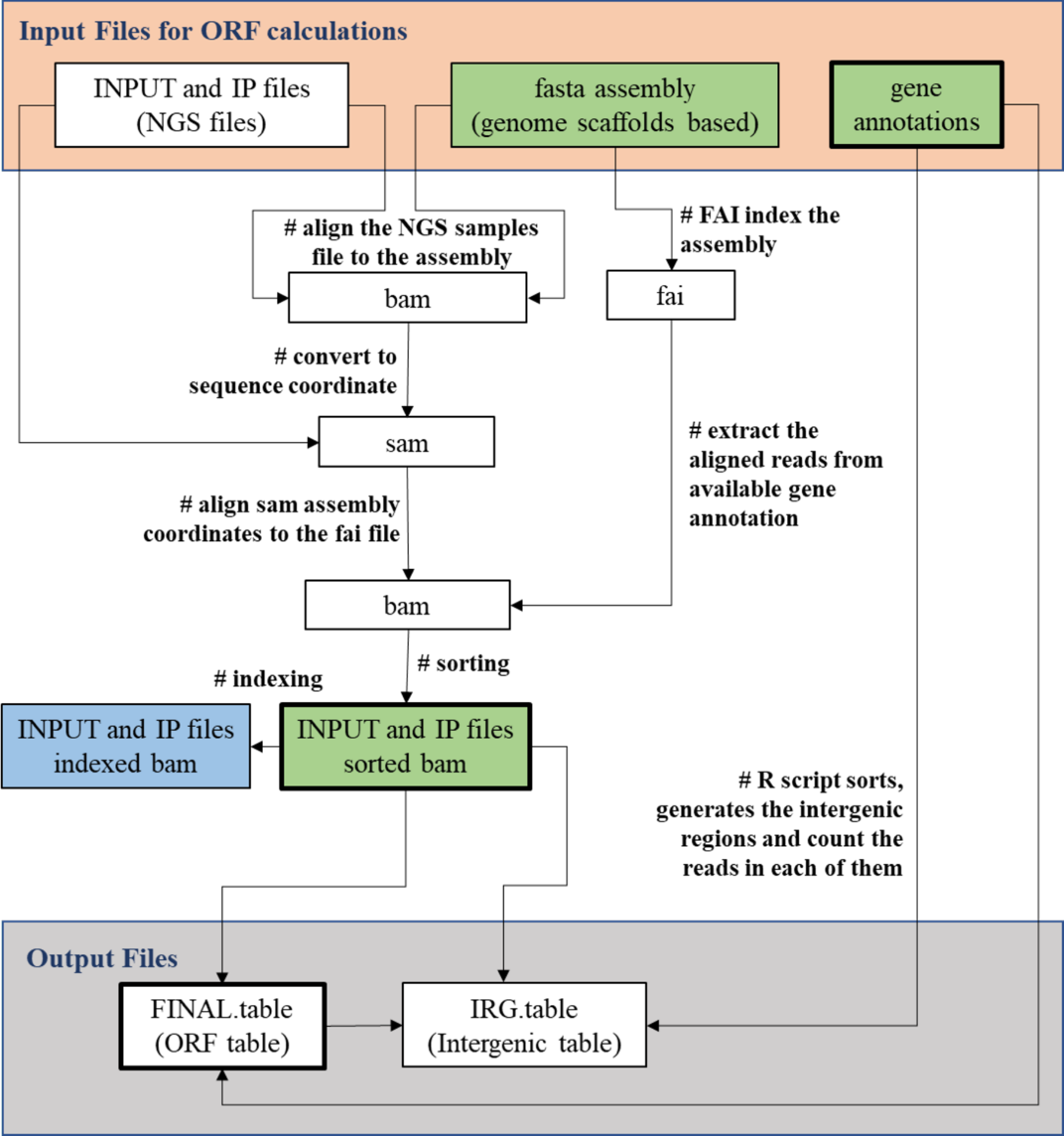}
       \caption{Core RACS pipeline overview. This flowchart represents the logic 
        steps implemented in the core pipeline. Boxes represent files and 
        file types as indicated in the text. Files with thick boxes represent 
        the Input Files for Intergenic calculations. Files in green are to 
        be uploaded to IGV. File in blue is needed for IGV but it does 
        not have to be uploaded to IGV. This file has to be kept 
        in the same folder directory than the sorted bam.}
        \label{fig:CorePipeline}
    \end{figure}
   \begin{figure}[h!]
	\includegraphics[width=\columnwidth]{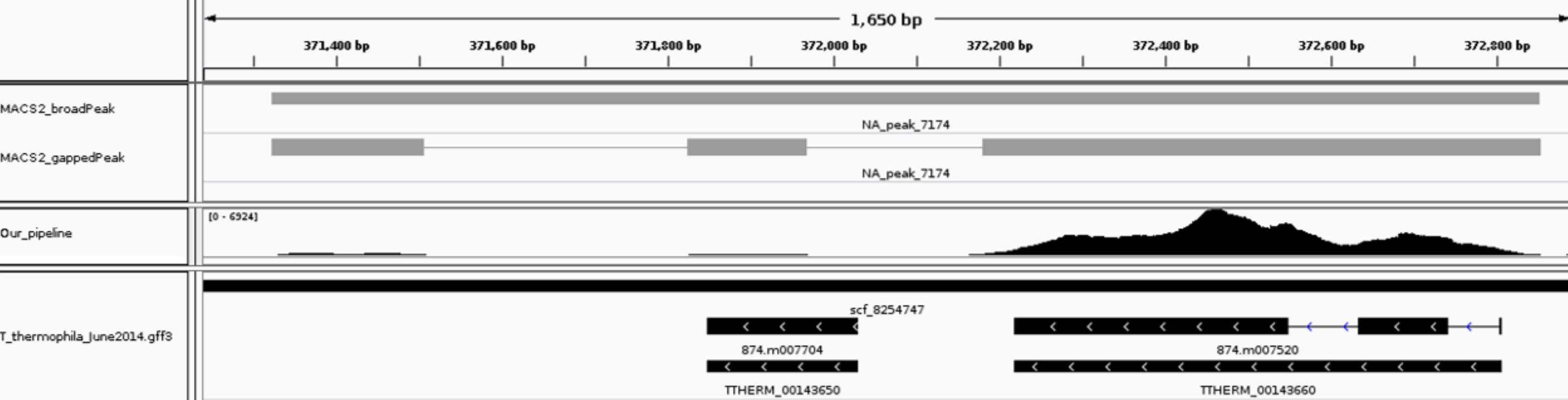}
       \caption{Visualization using IGV to compare MACS2 to our pipeline.
        Determination of an intergenic region obtained by MACS2 showing
        the broad and gapped peaks (first track) and by the pipeline
        presented in this paper (second track). Note that our pipeline
        shows graphical peaks  behaviour. The third track shows
        \textit{T.thermophila}'s genes.}
        \label{fig:genic_vs_MACS2}
\end{figure}

 \begin{figure}[h!]
	\includegraphics[width=\columnwidth]{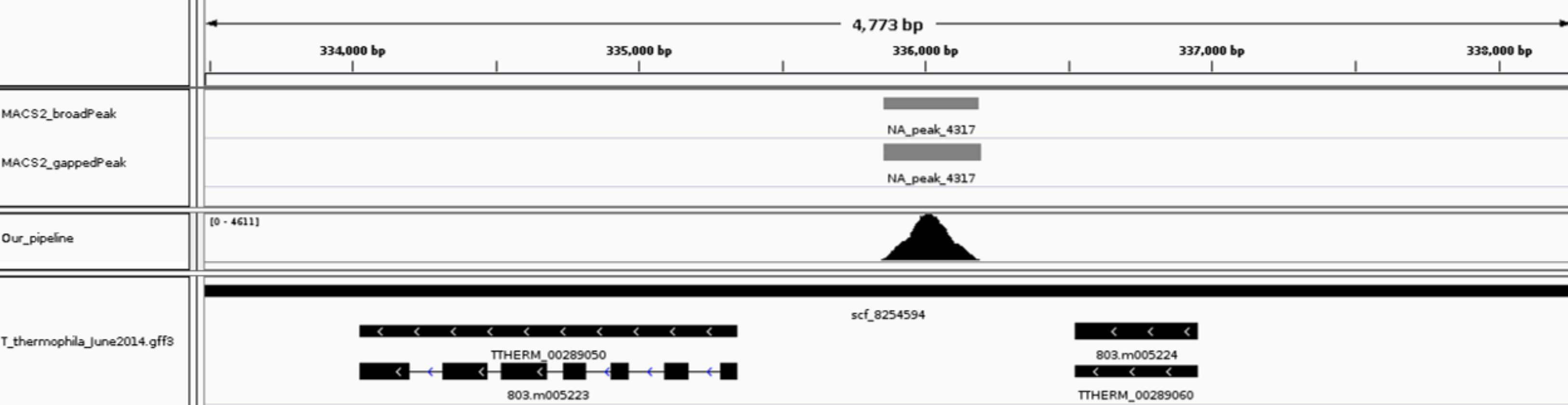}
       \caption{Results comparing the determination of the Genic Regions
    between the pipeline presented in this paper vs MACS2. Region obtained 
    by MACS2 (first track) and RACS pipeline (second track).
        MACS2 found two weak peaks that can be interpreted as 
        background by our pipeline.}
        \label{fig:intergenic_vs_MACS2}
    \end{figure}

   \begin{figure}[h!]
	\includegraphics[width=\columnwidth]{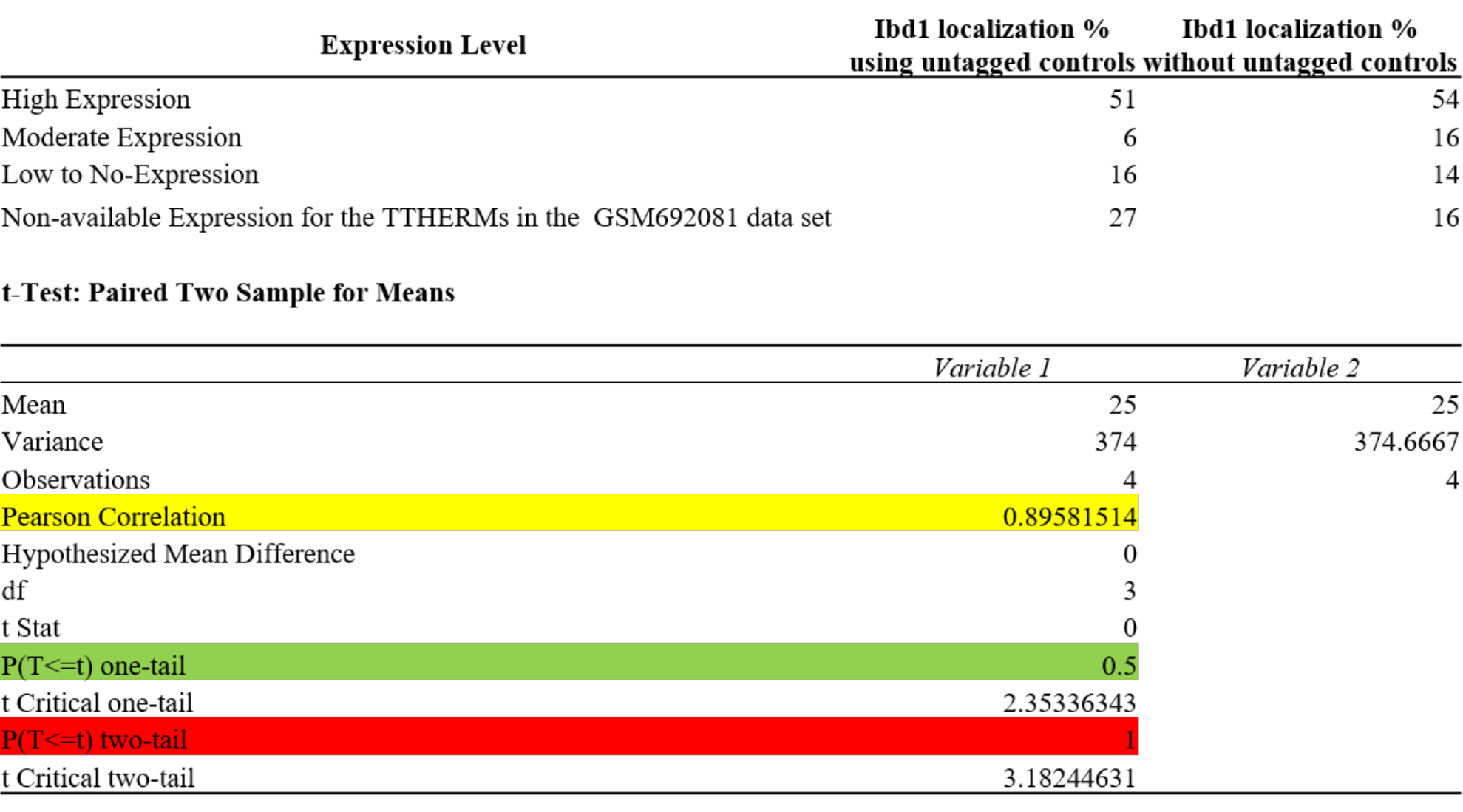}
        \caption{Comparison of Ibd1 localization presented in 
        \cite{Saettone2018} (without untagged controls) and 
        analyzed by RACS using untagged controls (current study). There is a 
        correlation of 0.896 and non-statistical differences between the 
        two data sets. 
        The data 
        presented in \cite{Saettone2018} uses an arbitrary cut-off. The data 
        presented in this paper does not use the arbitrary cut-off and instead 
        uses as cut-off the values obtained by the analyses of the untagged samples.}
        \label{fig:Table}
     \end{figure}

\begin{figure}[h!]
	\includegraphics[width=\columnwidth]{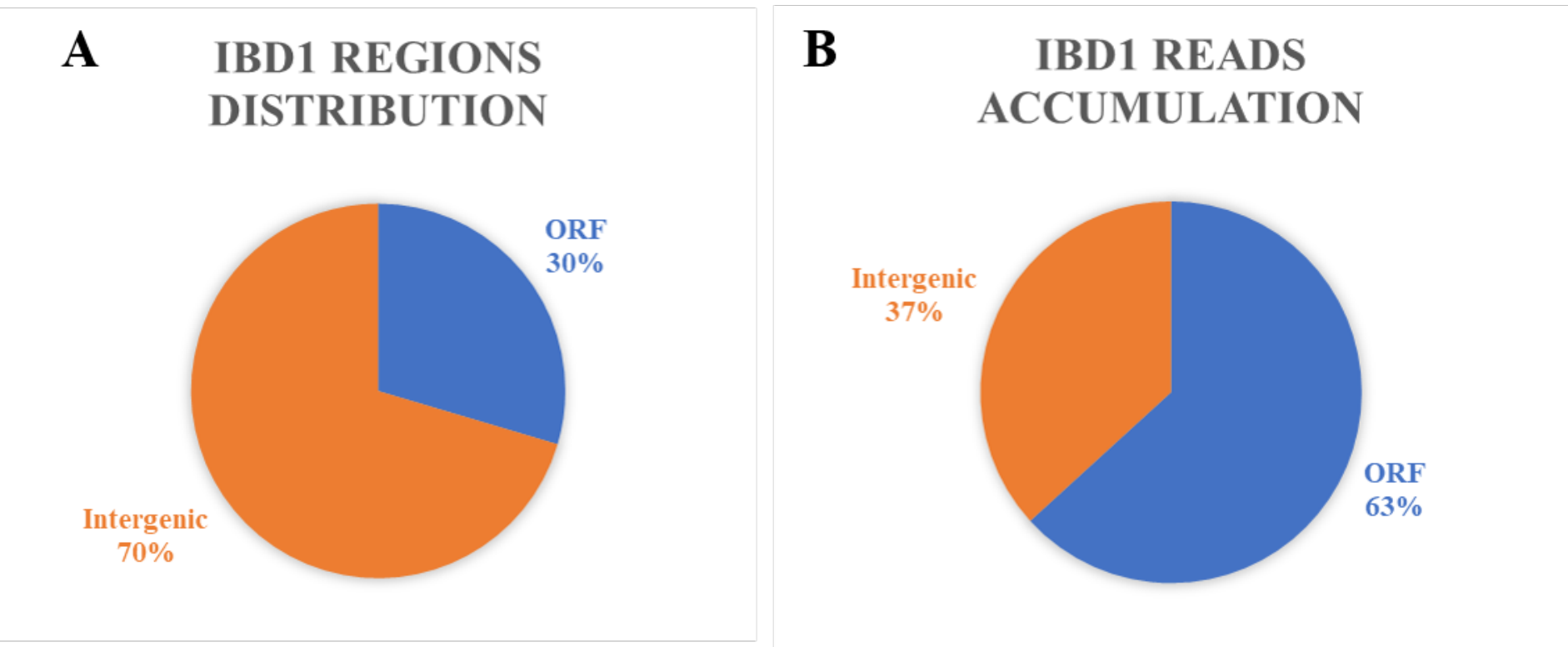}
	\caption{Ibd1 localization using RACS.
		\textbf{A.} Ibd1 localizes to more Intergenic regions than ORF.
		\textbf{B.} The majority of reads are found in ORF.
	These results take into consideration the updated information provided by the mock samples.}
        \label{fig:Ibd1_distribution}
    \end{figure}

\begin{figure}[h!]
	\includegraphics[width=\columnwidth]{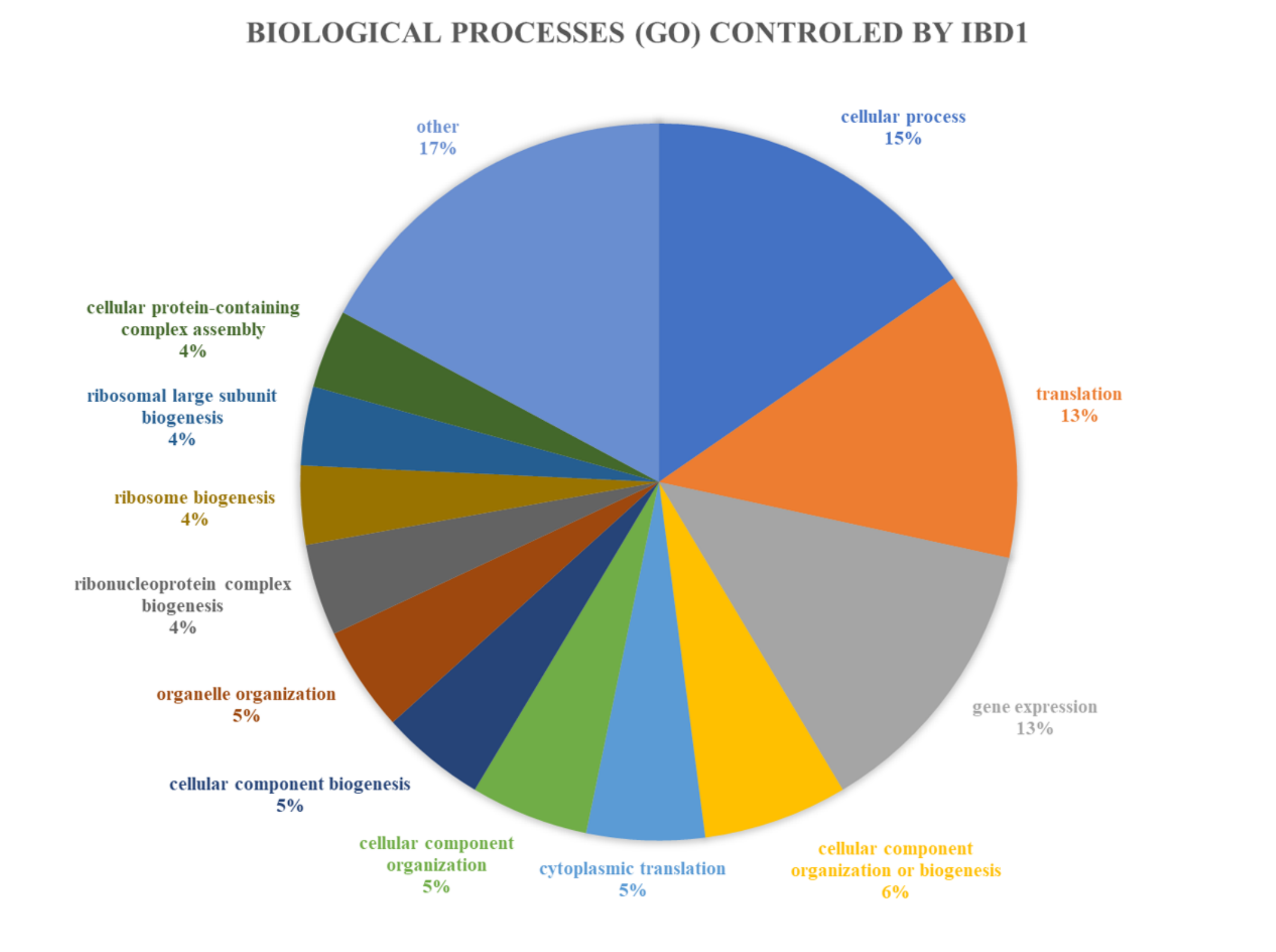}
	\caption{Gene Ontology (GO) analysis of genes controlled by Ibd1. }
        \label{fig:Ibd1_GO}
    \end{figure}

\clearpage
\appendix

\section{Abbreviations}
The following abbreviations are used in this manuscript:\\

\noindent
\begin{tabular}{@{}ll}
\\\\\\ChIP-Seq & Chromatin immunoprecipitation coupled to Next Generation Sequencing\\
NGS & Next Generation Sequencing\\
HPC & High Performance Computing\\
I/O & Input/Output \\
GPC & General Purpose Cluster \\
RACS & Rapid Analysis of ChIP-Seq data\\
MAC & Macronucleus\\
MIC & Micronucleus\\
ENCODE & Encyclopedia of DNA Elements\\
ORF & Open Region Frame \\
IGR & Intergenic Region \\
IP & Immunoprecipitated \\
FCS & Flowcell summary \\
N & Normalized
\end{tabular}

\end{backmatter}

\end{document}